\documentclass[12pt,preprint]{aastex}
\usepackage{lscape}

\def\ab{{\rm ab}}

\def\cc{{\rm c}}
\def\col{{\rm col}}
\def\disk{{\rm disk}}

\def\emm{{\rm em}}
\def\es{{\rm es}}
\def\g{{\rm g}}
\def\gr{{\rm GR}}
\def\in{{\rm in}}
\def\kk{{\rm K}}

\def\l{{\rm loc}}

\def\obs{{\rm obs}}
\def\orb{{\rm orb}}

\def\p{{\rm p}}
\def\pr{{\rm pr}}
\def\ss{{\rm s}}
\def\sgn{{\rm sign}}
\def\u{{\rm u}}
\def\vv{{\rm V}}
\def\w{{\rm w}}
\def\wp{{\rm WP}}

\begin{document}

\title{Inferring the Inclination of a Black Hole Accretion Disk
from Observations of its Polarized Continuum Radiation}

\author{Li-Xin Li\altaffilmark{1},
Ramesh Narayan\altaffilmark{2},
Jeffrey E. McClintock\altaffilmark{2}}

\altaffiltext{1}{Max-Planck-Institut f\"ur Astrophysik,                        
Karl-Schwarzschild-Str. 1, 85741 Garching, Germany}             

\altaffiltext{2}{Harvard-Smithsonian Center for Astrophysics, 60 Garden 
Street, Cambridge, MA 02138}

\begin{abstract}

Spin parameters of stellar-mass black holes in X-ray binaries are
currently being estimated by fitting the X-ray continuum spectra of
their accretion disk emission.  For this method, it is necessary to know
the inclination of the X-ray-producing inner region of the disk.  Since
the inner disk is expected to be oriented perpendicular to the spin axis
of the hole, the usual practice is to assume that the black hole spin is
aligned with the orbital angular momentum vector of the binary, and to
estimate the inclination of the latter from ellipsoidal modulations in
the light curve of the secondary star.  We show that the inclination of
the disk can be inferred directly if we have both spectral and
polarization information on the disk radiation.  The predicted degree of
polarization varies from 0\% to 5\% as the disk inclination changes from
face-on to edge-on.  With current X-ray polarimetric techniques the
polarization degree of a typical bright X-ray binary could be measured
to an accuracy of 0.1\% by observing the source for about 10 days.  Such
a measurement would constrain the disk inclination to within a degree or
two and would significantly improve the reliability of black hole spin
estimates.  In addition, it would provide new information on the tilt
between the black hole spin axis and the orbital rotation axis of the
binary, which would constrain any velocity kicks experienced by
stellar-mass black holes during their formation.

\end{abstract}


\section{Introduction}\label{intro}

X-ray polarimetry is an uncharted frontier with great promise for the
exploration of the behavior of accreting black holes and for the
measurement of black hole spin.  To date, only small X-ray polarimeters
have been flown and the only result of consequence has been the
measurement of the polarization of the Crab Nebula (Novick et al.\ 1972;
Weisskopf et al.\ 1978).  Anticipating the launch of larger and more
sensitive X-ray polarimeters in the future, this paper discusses
possible uses of polarization observations to study thermal emission
from accretion disks around stellar-mass black holes in X-ray binaries.

Polarization measurements at radio and optical wavelengths attest to the
power of polarimetry.  In radio astronomy, the measurement of
polarization is trivial.  However, its consequences have been of central
importance since Alf\'ven \& Herlofson (1950) first suggested the
synchrotron mechanism as the source of celestial radio emission.  For
example, studies of this strongly polarized radiation has allowed the
mapping of magnetic fields that are threaded throughout stellar coronae,
supernova shocks, relativistic jets, and the interstellar- and
intergalactic media.  At optical wavelengths, the measurement of
polarization is more difficult, but the payoffs have also been large.
For example, polarimetric observations of NGC~1068 provided a
spectacular confirmation of the unified model of AGN (Antonucci \&
Miller 1985).  Polarization measurements are most challenging at X-ray
wavelengths; however, sure instrumental approaches are known (\S6), and
the discovery space and potential rewards are large.

X-ray polarimetric information -- direction and degree -- increases the
parameter space used to investigate compact objects from the current two
-- spectra and time variability -- to four independent parameters that
models must satisfy (Rees 1975; Lightman \& Shapiro 1975; M\'esz\'aros
et al.\ 1988).  X-ray polarimetry can provide qualitatively new
information on source geometry and magnetic fields on spatial scales
comparable to a black hole event horizon.  In this paper, we follow the
pioneering work of \citet{sta77}, \citet{con77}, and \cite{con80}, and 
we narrow our focus to
consider solely the problem of polarized thermal emission generated by a
thin accretion disk around a stellar-mass black hole.  The theory
developed here of course applies as well to thermal accretion disks
around supermassive black holes.  We do not consider coronal effects
that can generate an X-ray reflection spectrum \citep[e.g.,][]{ros99,dov04}
and Comptonized components of emission \citep[e.g.,][]{gie99}.

The polarization model presented here extends our workhorse accretion
disk model {\sc kerrbb} (Li et al.\ 2005), which we have used to
estimate the spins of four stellar-mass black holes via fits to their
thermal continuum spectra (Shafee et al.\ 2006; McClintock et al.\ 2006;
Liu et al.\ 2008).  The extension of {\sc kerrbb} presented herein
includes a simple electron scattering atmosphere based on an
$\alpha$-viscosity prescription of the stress (Shakura \& Sunyaev 1973)
and a treatment of the polarized radiation field via the Stokes
parameters.  Radiative transfer effects will be considered in a future
paper.  

In \S\ref{motivation}, we discuss our motivations for this work,
focusing in particular on the application of polarization measurements
for estimating the spins of accreting black holes.  In \S\ref{polar},
we briefly review the polarization of disk emission induced by
electron scattering in an accretion disk atmosphere, assuming a
semi-infinite plane-parallel medium. In \S\ref{polar_remote}, we
present the mathematical scheme for calculating the polarization as
observed by a remote observer, after the polarized disk emission
propagates through the curved spacetime of a Kerr black hole. In
\S\ref{application}, we apply our polarization calculations to the
measurement of black hole spin, and in \S\ref{observation}, we briefly
discuss the prospects for observing the predicted polarization signal
from accreting stellar-mass black holes. Finally, in
\S\ref{conclusion}, we summarize the results.  Some mathematical
details related to the thin disk model we employ and the propogation
of the polarization vector in Kerr spacetime are presented in
Appendices \ref{disk_model} and \ref{p_prop}.

\section{Motivations}\label{motivation}

There are two principal and related motivations for undertaking this
work: (1) to validate a key assumption of the continuum-fitting method
of determining spin, and (2) to determine if the spins of stellar mass
black holes are aligned with the orbital plane.  We discuss these
topics in turn in the following subsections.  We then conclude this
section by commenting on three additional avenues for determining
black hole spin, viz., the Fe K line, high-frequency QPOs and X-ray
polarimetry.

\subsection{Validation of the Continuum-Fitting Method of Measuring Spin}

Using the continuum-fitting method, we have published spin estimates for
four stellar black holes (\S1), and we expect to publish spin estimates
for several more during the next few years.  In this work, it is
essential to use {\it thermal dominant} (formerly high soft state) data
(McClintock \& Remillard 2006).  A feature of this state is the
dominance of a soft blackbody-like component that is emitted by
optically-thick gas in the accretion disk.  A minor nonthermal tail
component of emission is also often present, but it contributes
typically only a few percent of the $\sim 1-10$ keV flux.  Thus, these
thermal-state spectra are largely free of the uncertain effects of
Comptonization and are believed to match closely the classic thin
accretion disk models of the early 1970s (Shakura \& Sunyaev 1973;
Novikov \& Thorne 1973).  There is an abundance of such suitable
thermal-state continuum spectra available in the NASA HEASARC archives
for most black hole binaries.  These spectra are ideally suited for
study using the polarization model presented in this paper.

The promise of the continuum-fitting method for determining black hole
spin is attested to by 20 years of experience that has demonstrated
for the thermal dominant state the presence of a stable inner disk
radius (see \S4 in McClintock et al.\ 2008).  The inner edge is
presumably closely tied to the innermost stable circular orbit (ISCO).
X-ray observations, coupled with ground-based optical and infrared
observations, are used to measure the radius of this orbit.  Applying
the continuum-fitting method is analogous to the familiar problem of
determining the radius of a star of known distance given its effective
temperature and flux (McClintock et al.\ 2008).  Accordingly, this
method requires knowledge of the luminosity of the X-ray source, which
depends not only on its distance but also on the projected area and
hence inclination of the accretion disk.  Accurate values of the
inclination of the orbital plane $i_{\rm orb}$, distance $D$, and
black hole mass $M$ are readily obtained from ground-based
observations.  This optical work has been underway for 35 years and is
now undergoing a renaissance \citep[e.g.,][]{oro07}.

The continuum-fitting method is straightforward to apply.  All the
required data are readily obtainable, and even the theory of disk
accretion in strong gravity is tractable (Narayan et al.\ 2008; Shafee
et al. 2008b).  However, the reliability of the continuum-fitting
method is called into question by a single assumption, viz., that the
plane of the inner X-ray-emitting portion of the disk is aligned with
the binary orbital plane, whose inclination angle $i_{\rm orb}$ is
determined from optical observations.  Unfortunately, the
continuum-fitting method cannot fit for the inclination of the inner
disk and check for disk warp (\S2.2) because there is a degeneracy
between the inclination and spin parameters (\S~5.2; see also Liu et
al.\ 2008).  As we show, polarimetry observations can easily determine
the inclination of the inner disk.  This is our principal motivation
for writing this paper.

\subsection{The Inclination of the Inner Accretion Disk}\label{inclination}

Whether a black hole's spin is aligned with its accretion disk and the
time scale for any misalignment to dissipate are topics that have
received considerable attention recently (Fragile \& Anninos 2005; King
et al.\ 2005; Lodato \& Pringle 2006; Fragile et al.\ 2007; Martin et
al.\ 2007).  In this paper, we restrict our attention to the question of
alignment in the case of stellar black holes in X-ray binaries.  If a
black hole's spin were to be misaligned from the orbital vector, the
inner disk would be warped away from the outer disk by Lense-Thirring
precession and the Bardeen-Petterson effect \citep{bar75}, which would 
be expected to
cause a global precession of the main body of the disk and have
important observational consequences (e.g., enhanced mass accretion rate
and jet precession; Fragile et al.\ 2007; Martin et al.\ 2007).

Are significant misalignments expected in the case of accreting
stellar black holes?  On the one hand, it seems plausible that the
spin will be aligned with the orbit vector given the strong tidal
forces acting on the pre-supernova star (Zahn 1977) and, as generally
assumed (e.g., Fryer \& Kalogera 2001), that collapse of the stellar
core to a black hole will be nearly spherical and mostly free of the
strong shocks and velocity kicks that nascent neutron stars suffer.
Even if shocks and kicks do occur, it is expected that their effects
will be less for black holes by the factor $M_{\rm BH}/M_{\rm NS} \sim
7$ (e.g., Fryer \& Young 2007).  On the other hand, there is some
evidence for warped disks based on radio-jet data (Maccarone 2002),
although this evidence is weak (Narayan \& McClintock 2005).  One can
attempt to infer inclination via studies of radio jets (Mirabel \&
Rodr\'iguez 1999) and Fe K emission line profiles (Fabian et al.\
2000).  However, as we show in this paper, the direct and sure
approach is by measuring the degree of polarization of the thermal
accretion-disk radiation, which depends strongly on the inclination
angle $i_{\rm disk}$ of the inner disk.

\subsection{Three Additional Methods of Measuring Black Hole Spin}

The Fe K-line method of determining spin is applicable to supermassive
black holes as well as stellar black holes (Miller 2007; Reynolds \&
Fabian 2008).  It has received considerable attention and has recently
yielded quantitative estimates of spin for the Seyfert 1.2 galaxy
MCG-6-30-15 (Brenneman \& Reynolds 2006) and for the stellar black hole
GX 339-4 (Miller et al.\ 2008). 

High-frequency QPOs may develop into a precise method of determining
spin once the correct model of the oscillations has been identified
(e.g., Wagoner et al.\ 2001; T\"or\"ok et al.\ 2005; Schnittman 2005).
X-ray polarimetry may also provide a means of determining spin.  This
possibility is implicit in the work of Connors et al.\ (1980), which
shows that for reasonable assumptions one expects the plane of
polarization to swing smoothly through a large range of angle as the
photon energy increases.  By modeling this effect and the degree of
polarization one can hope to obtain an independent estimate of black
hole spin, a topic that we touch upon in \S5.1.

Thus, there are potentially a total of four methods of measuring black
hole spin, two of which -- the continuum-fitting method and the Fe
K-line method -- are presently delivering results.  Because black hole
spin is such a fundamental parameter, it is important to attempt to
measure it by as many of these methods as possible, as this will
provide arguably the best possible check on the results.

\section{Polarization of Disk Emission Induced by Electron Scattering}
\label{polar}

We discuss in this and the following section the technical details of
how we calculate the polarization of the thermal disk emission.  We
present additional details in the Appendices.  The presentation is
pedagogical so that this paper and Li et al. (2005) together provide a
complete and self-contained description of the calculations.

The radiation generated deep inside an accretion disk is initially
unpolarized, but it becomes polarized as a result of electron
scattering in the disk atmosphere.  For an optically thick and
geometrically thin accretion disk, the disk atmosphere is locally well
described by a semi-infinite plane-parallel medium.  We make this
approximation.

A radiation field is completely described by four Stokes parameters:
$I$ ($\equiv I_l+I_r$, the total intensity), $Q$ ($\equiv I_l-I_r$),
$U$, and $V$ \citep{cha60}.  Here the subscripts $l$ and $r$ refer,
respectively, to polarization in the meridian plane (i.e., the plane
containing the symmetry axis and the line-of-sight to the observer)
and at right angles to it.  In a plane-parallel atmosphere with no
incident radiation, the axial symmetry of the radiation field requires
that the plane of polarization be either along $l$ or $r$.  Therefore,
$U=V=0$, and the two parameters $I$ and $Q$, or equivalently the two
intensities $I_l$ and $I_r$, suffice to characterize the radiation.
The degree of polarization of the radiation field is then given by
\begin{eqnarray}
	P \equiv \frac{1}{I}\left(Q^2+U^2+V^2\right)^{1/2} 
       	       	= \frac{\left|I_r-I_l\right|}{I_r+I_l} \;.   
	\label{P}
\end{eqnarray}

At the surface of a pure scattering atmosphere, the two intensities
$I_l$ and $I_r$ are determined by
\begin{eqnarray}
	I_l(\mu) = \frac{3}{8\sqrt{2}\pi} FH_l(\mu) q \;, \hspace{1cm}
	I_r(\mu) = \frac{3}{8\sqrt{2}\pi} FH_r(\mu) (\mu+c) \;,
	\label{I_sol}
\end{eqnarray}
where $\mu\equiv\cos\theta$, $\theta$ is the polar angle of the
direction of propagation of the photon to the disk normal, $F$ is the
total emitted flux density, $H_l$ and $H_r$ are Chandrasekhar's
$H$-functions \citep{cha60}, and $q$ and $c$ are two integral
constants.  Following \citet{cha60}, we have suppressed the
dependences of the relations on the photon energy since no ambiguity
is likely to arise.  The constants $q$ and $c$ are given by
\citep[{\S}X]{cha60}
\begin{eqnarray}
	q = \frac{8(A_1+2\alpha_1)-6(A_0\alpha_1 +\alpha_0A_1)}
		{3\left(A_1^2+2\alpha_1^2\right)} \;, \hspace{1cm}
	c=\left(1-\frac{q^2}{2}\right)^{1/2} \;,
\end{eqnarray}
where $\alpha_n$ and $A_n$ are the moments of order $n$ of $H_l(\mu)$ and
$H_r(\mu)$:
\begin{eqnarray}
	\alpha_n \equiv \int_0^1 H_l(\mu)\mu^n d\mu \;, \hspace{1cm}
		A_n \equiv \int_0^1 H_r(\mu)\mu^n d\mu \hspace{1cm}
		(n\ge 0) \;.
	\label{h_mom}
\end{eqnarray}

\begin{figure}
\epsscale{0.9}
\plotone{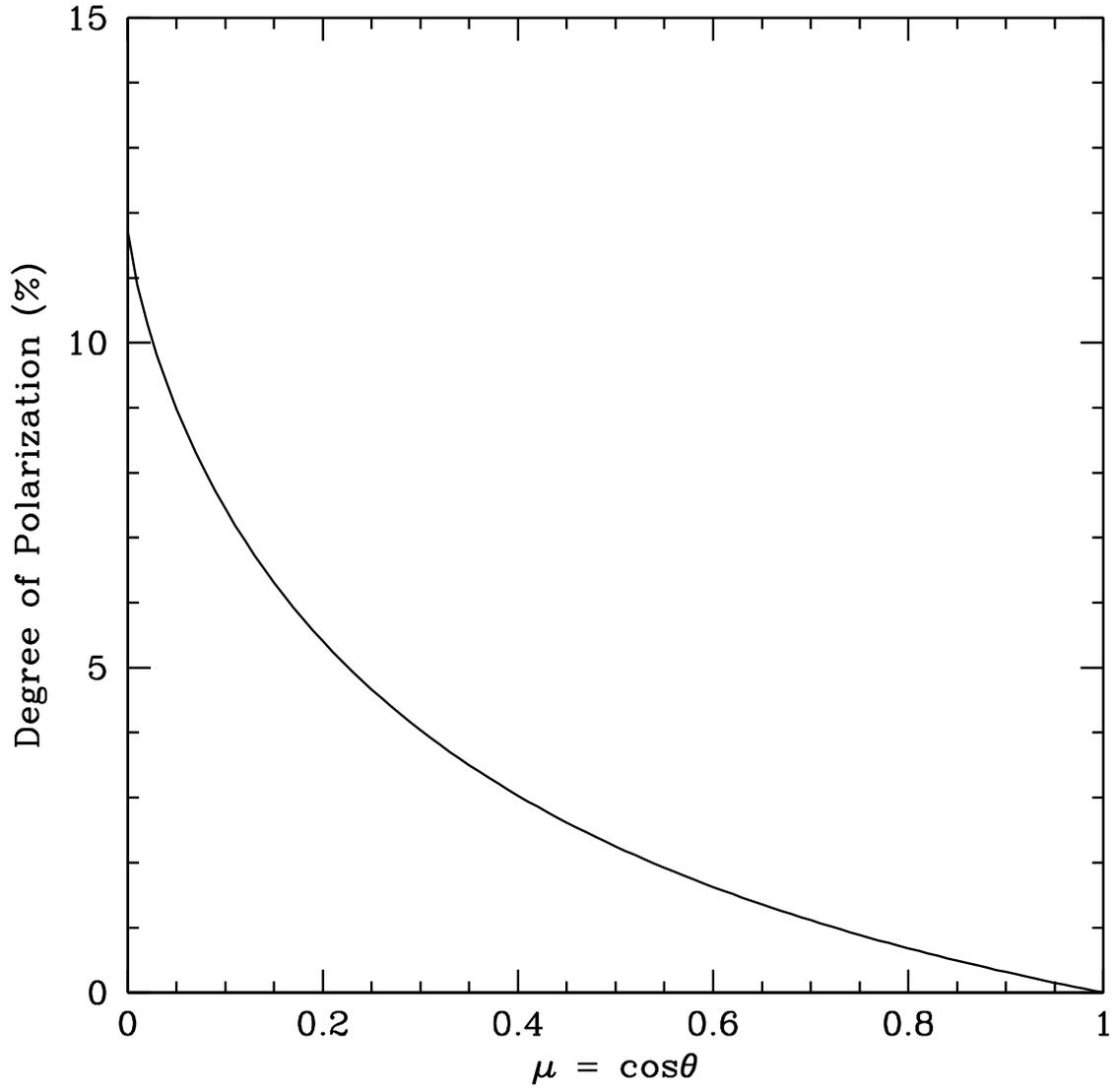}
\vspace{0.8cm}
\caption{Degree of polarization as a function of $\mu = \cos\theta$
for a semi-infinite scattering atmosphere, where $\theta$ is the angle
between the line-of-sight and the normal to the plane of the
atmosphere.
\label{p_deg_cha}}
\end{figure}

Figure~\ref{p_deg_cha} shows the degree of polarization $P$ as a
function of $\mu$ for a scattering atmosphere. At $\mu=1$ ($\theta =
0$), we have $P = 0$ because of symmetry. As $\mu$ decreases (i.e.,
$\theta$ increases), $P$ increases monotonically and reaches its
maximum value of $11.7\%$ at $\mu=0$ ($\theta = \pi/2$).  For a
semi-infinite atmosphere we always have $I_r(\mu)>I_l(\mu)$ \citep[see
table XXIV in][]{cha60}, except at $\mu=1$ ($\theta=0$) where
$I_r(\mu)=I_l(0,\mu)$.  Hence we have $\psi=\pi/2$ for all $\mu$,
where $\psi$ is the angle of the linear polarization vector relative
to the meridian plane.

The above results are for a pure scattering semi-infinite
atmosphere. However, an accretion disk also has absorption processes
which may be important, especially at large disk radii where the
temperature is low.  Absorption tends to destroy the polarization of
photons and hence to reduce the degree of polarization.  We defer a
detailed computation of the effect of absorption to a later paper.
Here, following \citet{lao90} and \citet{che91}, we simply reduce the 
degree of polarization by the following approximate factor
\begin{eqnarray} 
	q_\w = \frac{\tau_\es}{\tau_\es+\tau_{\ab}} \;, \label{q_parameter}
\end{eqnarray}
where $\tau_\es$ is the optical depth due to electron scattering, and
$\tau_\ab$ is the optical depth due to photon absorption.  Thus, we write
the degree of polarization of photons emerging from the disk surface as
\begin{eqnarray}
	P_\emm = q_\w P_{\emm,0} \;, \label{P_emm}
\end{eqnarray}
where $P_{\emm,0}$ is the degree of polarization for a pure electron
scattering atmosphere and is given by equations (\ref{P}) and
(\ref{I_sol}).  We assume that the orientation of the polarization is
not affected by absorption.

To complete the description of the polarized emission from an
accretion disk, we need the flux $F$ and the optical depths $\tau_\es$
and $\tau_\ab$ as functions of the radius $R$.  These are obtained
using the disk model described in Appendix \ref{disk_model}.

\section{Polarization of the Disk Emission as Observed by a Remote Observer}
\label{polar_remote}

An arbitrary radiation field can be decomposed into an unpolarized
component, with Stokes parameters
\begin{eqnarray}
	\left\{(1-P) I,\; 0,\; 0,\; 0\right\} \;,
\end{eqnarray}
and a completely polarized component, with Stokes parameters
\begin{eqnarray}
	\left\{P I,\; Q,\; U,\; V\right\} \;,
\end{eqnarray}
where $P$ is the degree of polarization.  Further we have
\begin{eqnarray}
	\tan 2\psi = \frac{U}{Q} \;, \hspace{1cm} \sin 2\beta =
		\frac{V}{(Q^2+U^2+V^2)^{1/2}} \;,
	\label{angle}
\end{eqnarray}
where $\psi$ is the angle between the major axis of the polarization
ellipse and the direction described by the subscript $l$ (see
\S\ref{polar}),\footnote{We denote the angle of the plane of
polarization from the direction $l$ by $\psi$ \citep[rather than the
$\chi$ in][]{cha60} to be consistent with \citet{con80}.} and
$\tan\beta$ is the ratio of the minor axis to the major axis of the
polarization ellipse.  Photons emitted by the disk are linearly
polarized, so we have $V=0$ and $\beta=0$. Then, by the definition of
the degree of polarization $P$, we have
\begin{eqnarray}
	Q = I_\p \cos 2\psi \;, \hspace{1cm} U = I_\p \sin 2\psi \;,
        \hspace{1cm} I_\p \equiv PI \;.
\end{eqnarray}
Equivalently, we can write
\begin{eqnarray}
	Q + i U = I_\p e^{2i\psi} \;. \label{QiU}
\end{eqnarray}

When several independent streams of light are combined, the Stokes
parameters for the combined radiation are the sums of the respective
Stokes parameters of the individual streams \citep{cha60}; we refer to
this as the {\em superposition theorem}. Therefore, we may consider
the unpolarized and polarized components of the radiation separately.

For the {\em polarized component}, consider a beam of perfectly
polarized radiation emitted by an infinitesimal surface element on the
disk and received by an observer along an infinitesimal solid angle
element $d\Omega_\obs$.  Let this radiation have intensity
$I_{\p,\obs}$ and polarization angle $\psi_\obs$ as measured by the
observer. Then, the observed Stokes parameters are
\begin{eqnarray}
	\left\{I_{\p,\obs},\; Q_\obs,\; U_\obs,\; 0\right\} \;,
\end{eqnarray}
where
\begin{eqnarray}
	Q_\obs = I_{\p,\obs}\cos 2\psi_\obs \;, \hspace{1cm}
		U_\obs = I_{\p,\obs}\sin 2\psi_\obs \;.
\end{eqnarray}
Summing over the radiation received from all disk elements we have, by
the superposition theorem,
\begin{eqnarray}
	\langle I_{\p,\obs}\rangle = \frac{1}{\Delta\Omega_\obs} 
		\int I_{\p,\obs} d\Omega_\obs \;,
	\label{I_p}
\end{eqnarray}
and
\begin{eqnarray}
	\langle Q_\obs\rangle + i \langle U_\obs\rangle &=& 
		\frac{1}{\Delta\Omega_\obs} \int \left(Q_\obs + i 
		U_\obs\right) d\Omega_\obs \nonumber\\
		&=& \frac{1}{\Delta\Omega_\obs} \int I_{\p,\obs} 
		e^{2i\psi_\obs} d\Omega_\obs\;,
	\label{QU_obs}
\end{eqnarray}
where `$\langle\, \rangle$' denotes an average over solid angle, and
$\Delta\Omega_\obs$ is the total solid angle subtended by the disk on
the sky.

Note that, in general, we have
\begin{eqnarray}
	\langle Q_\obs\rangle^2 + \langle U_\obs\rangle^2 < \langle 
		I_{\p,\obs}\rangle^2 \;,
\end{eqnarray}
i.e., the radiation is not fully polarized.  Even though the
individual contributions from each disk element may be perfectly
polarized, the radiation received by the observer can still become
partially polarized through averaging because the individual rays may
have different values of $\psi_\obs$ at the observer.  In the context
of an accretion disk around a black hole, even if all the emitted rays
have the same polarization direction at their points of emission,
geodesic propagation can cause changes in $\psi$ along each photon
trajectory and cause a reduction in the observed degree of
polarization.
In other words, the rotation of the disk and the spin of the black
hole can alter the polarization state of the original radiation and
destroy the observed polarization at some level.  (In contrast,
whatever unpolarized radiation is emitted by the disk remains
unpolarized at the observer; thus, geodesic propagation and averaging
can only decreased the degree of polarization.)

For the {\em unpolarized component}, we have
\begin{eqnarray}
	\langle I_{\u,\obs}\rangle = \frac{1}{\Delta\Omega_\obs} 
		\int I_{\u,\obs} d\Omega_\obs \;,
	\label{I_u}
\end{eqnarray}
where $I_{\u,\obs} \equiv I_{\obs} - I_{\p,\obs}$.  By equations
(\ref{I_p}) and (\ref{I_u}) we have
\begin{eqnarray}
	\langle I_{\u,\obs}\rangle + \langle 
		I_{\p,\obs}\rangle = \frac{1}{\Delta\Omega_\obs} 
		\int \left(I_{\u,\obs} + I_{\p,\obs}\right) d\Omega_\obs 
		= \frac{1}{\Delta\Omega_\obs} \int I_{\obs} d\Omega_\obs 
		= \langle I_{\obs}\rangle \;,
	\label{I_tot}
\end{eqnarray}
i.e., the total intensity is conserved.

Using the fact that $I_{E_\l}/E_\l^3$ is invariant along the path of a
photon, where $E_\l$ is the photon energy measured by a local observer
along the photon path \citep{mis73}, equation (\ref{QU_obs}) can be
rewritten as \citep[c.f. eq. 9 of][]{con80}
\begin{eqnarray}
	\langle Q_\obs\rangle + i \langle U_\obs\rangle = \frac{1}
                {\Delta\Omega_\obs} \int g^3 P_{\emm} I_{\emm} 
		e^{2i\psi_\obs} d\Omega_\obs \;,
	\label{QU_obs2}
\end{eqnarray}
where $P_{\emm}$ is the degree of polarization of the radiation at the
time that the radiation emerges from the disk surface, and $g$ is the
redshift factor of the photon \citep{li05}.  Similarly, we can rewrite
the total intensity as\footnote{Note that, in order to simplify the
notation, we have suppressed the dependenc on $E_{\emm}$ -- the photon
energy as measured by a local observer corotating with the disk -- in
our expressions.  This should cause no ambiguity.}
\begin{eqnarray}
	\langle I_{\obs}\rangle = \frac{1}{\Delta\Omega_\obs} \int g^3 
		I_{\emm} d\Omega_{\rm obs} \;.
	\label{I_tot2}
\end{eqnarray}
The observed average degree of polarization is then given by
\begin{eqnarray}
	\langle P\rangle = \frac{1}{\langle I_{\obs}\rangle}{\sqrt{
		\langle Q_\obs\rangle^2 + \langle U_\obs\rangle^2}} \;,
	\label{P_obs}
\end{eqnarray}
while the observed average angle of polarization, $\langle \psi\rangle$, is 
determined by\footnote{We do not use $\tan (2\langle \psi\rangle) = \langle 
U_\obs\rangle/\langle Q_\obs\rangle$. This single equation cannot determine
the value of $\langle\psi\rangle$, since $\tan(2\langle \psi\rangle)$ is
unchanged under the transformation $\langle Q_\obs\rangle\rightarrow 
-\langle Q_\obs\rangle$ and $\langle U_\obs\rangle\rightarrow -\langle 
U_\obs\rangle$. This is related to the fact that the primitive period of 
$\sin x$ (and $\cos x$) is $2\pi$ while the primitive period of $\tan x$ is 
$\pi$.}
\begin{eqnarray}
	\sin (2\langle \psi\rangle) = \frac{\langle U_\obs\rangle}
		{\sqrt{\langle Q_\obs\rangle^2 + \langle U_\obs\rangle^2}} \;,
	\hspace{1cm}
	\cos (2\langle \psi\rangle) = \frac{\langle Q_\obs\rangle}
		{\sqrt{\langle Q_\obs\rangle^2 + \langle U_\obs\rangle^2}} \;.
	\label{psi_obs}
\end{eqnarray}
In the definitions of the above dimensionless quantities the total
solid angle $\Delta\Omega_\obs$ cancels out.

The solution to equation (\ref{psi_obs}) is 
\begin{eqnarray}
	\langle\psi\rangle = \langle\psi\rangle_\pr + n\pi \;,
	\label{psi_obs2}
\end{eqnarray}
where $n$ is any integer, and the primitive angle $\langle\psi\rangle_\pr$
(defined to be in the range $0$--$\pi$) is given by
\begin{eqnarray}
	\langle\psi\rangle_\pr = \left\{\begin{array}{ll}
		\frac{1}{2}\arccos\xi_Q \;, & \left(\langle U_\obs\rangle>0; 
                0<\langle\psi\rangle_\pr<\frac{\pi}{2}\right) \\
		\pi - \frac{1}{2}\arccos\xi_Q\;, & \left(\langle U_\obs
                \rangle<0; \frac{\pi}{2}<\langle\psi\rangle_\pr<\pi\right)
                \end{array}\right. \;,
\end{eqnarray}
where
\begin{eqnarray}
	\xi_Q \equiv \frac{\langle Q_\obs\rangle}{\sqrt{\langle Q_\obs
		\rangle^2 + \langle U_\obs\rangle^2}} \;.
        \label{xi_QU}
\end{eqnarray}
In equation (\ref{QU_obs2}), the degree of polarization $P_\emm$ is
evaluated at the disk surface by equation (\ref{P_emm}), but the angle
of polarization $\psi_\obs$ is evaluated at the observer and is
related to the angle at the disk surface by the propagation equation
of the polarization vector.  The propagation equation is described in
detail in Appendix \ref{p_prop}.

In the limit of a semi-infinite plane atmosphere, the polarization
vector of the emerging radiation is in the disk plane, so we have
$\psi_\emm = \pi/2$ (see \S\ref{polar}). Since $\psi_\obs$ appears in
equation (\ref{QU_obs2}) in the form of $\cos 2\psi_\obs$ and $\sin
2\psi_\obs$, addition of $n\pi$ to $\psi_\obs$ ($n=\pm 1, \pm 2,...$)
does not change the final results. Hence, at the observer, by equation
(\ref{psi_inf_sol}) we can write
\begin{eqnarray}
	\psi_\obs = \frac{1}{2}\pi - \Phi_\gr \;,   
        \label{psi_obs_inf}
\end{eqnarray}
where $\Phi_\gr$ is calculated by equations (\ref{psi_gr1}),
(\ref{psi_gr2}), and (\ref{psi_gr_sol}) as described in Appendix
\ref{p_prop}.

Relativistic effects on photon polarization are reflected in the above
formula in two apsects. First, relativistic effects lead to a rotation to
the plane of polarization by an angle $\Phi_\gr$. Second, because of light
bending the angle of the photon wave vector from the disk normal as the 
photon leaves the disk surface, $\theta$, differs from the disk inclination 
angle, $i_\disk$. The Newtionian limit is obtained by setting $\Phi_\gr
=0$ and $\theta=i_\disk$. By equation (\ref{psi_obs_inf}) we then have
$\psi_\obs = \pi/2$. Then, by equation (\ref{QU_obs2}) we have $\langle
U_\obs\rangle = 0$ and
\begin{eqnarray}
	\langle Q_\obs\rangle = -\frac{P_{\emm,0}}{\Delta\Omega_\obs} 
                 \int q_\w g^3 I_{\emm}d\Omega_\obs \;,
	\label{QU_obs2a}
\end{eqnarray}
where equation (\ref{P_emm}) has been used, and $P_{\emm,0} = P_{\emm,0}
(\theta = i_\disk)$. Hence, in the Newtonian limit we have $\langle\psi
\rangle = \pi/2$, and
\begin{eqnarray}
	\langle P\rangle = P_{\emm,0} \frac{\int q_\w g^3 I_{\emm}d
                 \Omega_\obs}{\int g^3 I_{\emm}d\Omega_\obs} \;.
	\label{P_obs_New}
\end{eqnarray}
The high energy spectrum is dominated by photons emitted by the inner disk
region where photon absorption is not important ($q_\w \approx 1$), hence 
we have $\langle P\rangle \approx P_{\emm,0}$ for high energy photons.

As we will see in the next section, relativistic effects make the observed
features of photons emitted by the disk dramatically different from that
predicted by the Newtonian theory.

\section{Applications to the Measurement of Black Hole Spin}
\label{application}

\subsection{Comparison with Connors, Piran \& Stark (1980)}

Since the present work is a modern update on the pioneering work of
Connors et al. (1980), we begin by recomputing the models discussed in
their paper.  Appendix \ref{p_prop} describes in detail how we
calculate the polarization angle of an infinitesimal photon stream as
observed by an observer at infinity.  A key quantity is the primitive
rotation angle $\Phi_\gr$ defined by equations (\ref{psi_gr1}),
(\ref{psi_gr2}), and (\ref{psi_gr_sol}).  The calculation of
$\Phi_\gr$ requires an evaluation of the dimensionless quantities
$\tilde{X}$, $\tilde{Y}$, $\tilde{S}$, and $\tilde{T}$
(eqs. \ref{X_til}--\ref{ct2}), in addition to the orientation of the
photon wave vector relative to the disk plane ($\theta$ and $\phi$),
the photon redshift factor $g$, and the Lorentz factor $\Gamma$ of the
disk fluid. All these quantities can be evaluated with the formulae
given in Appendix \ref{p_prop} of this paper and Appendix C of
\citet{li05}.

The polarization angle $\psi_\obs$ is related to the rotation angle
$\Phi_\gr$ by equation (\ref{psi_obs_inf}). After calculating the
angle $\psi_\obs$ for each light ray that reaches the observer within
an infinitesimal solid angle $d\Omega_\obs$, one can evaluate the
integrals in equations (\ref{QU_obs2}) and (\ref{I_tot2}) with the
ray-tracing technique described in \citet{li05}. Then the net degree
of polarization $\langle P\rangle$ and angle of polarization $\langle
\psi\rangle$ of the combined radiation, as seen by the observer, can
be calculated via equations (\ref{P_obs}) and
(\ref{psi_obs2})--(\ref{xi_QU}).  In the calculations reported here,
the effect of photon absorption has been taken into account with the
simple approach outlined in \S\ref{polar}. That is, photon absorption
reduces the degree of polarization by introducing a multiplying factor
$q_\w$ defined in equation (\ref{q_parameter}).

\begin{figure}
\epsscale{0.9}
\plotone{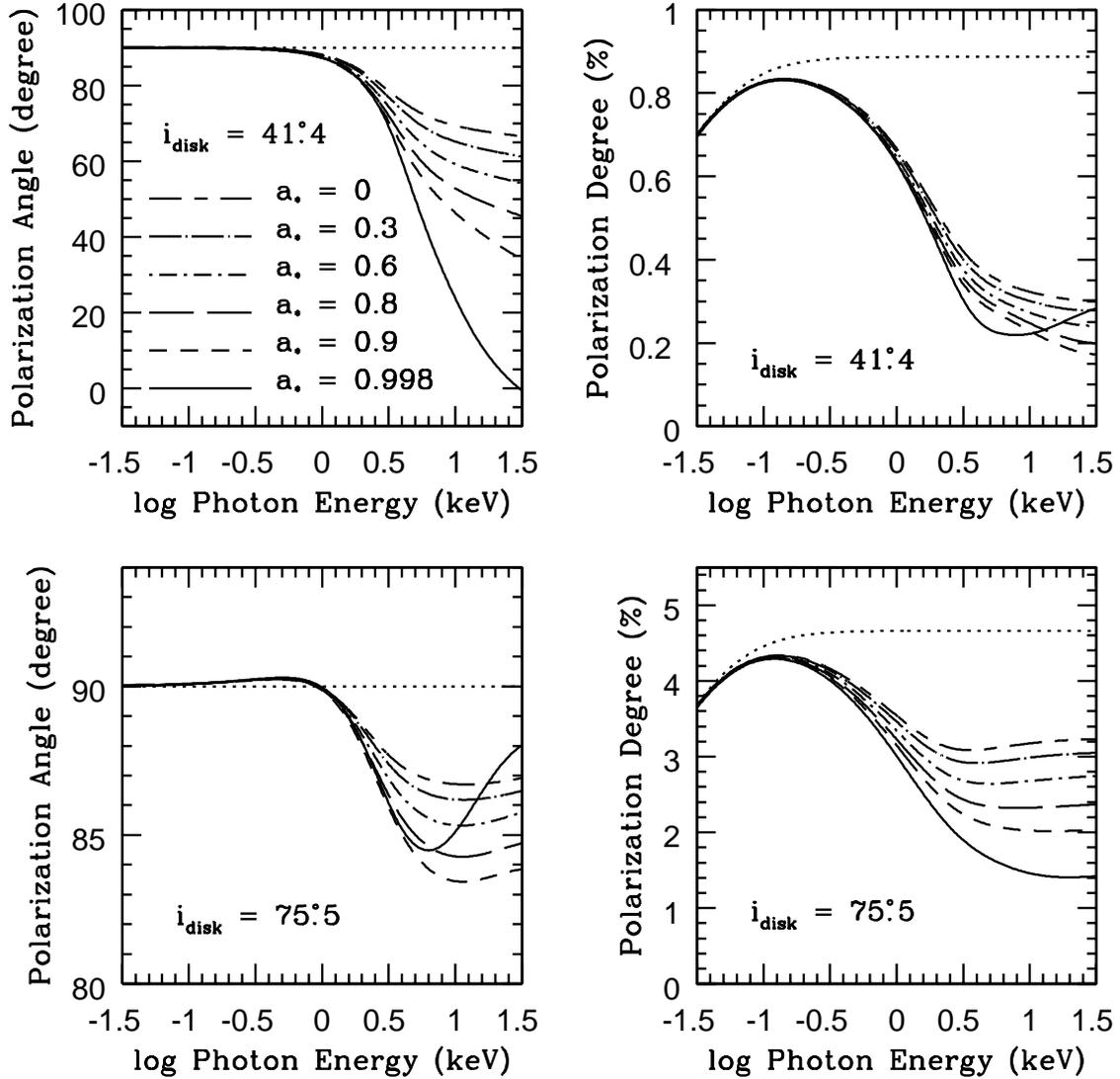}
\vspace{0.8cm}
\caption{Polarization angle and degree of polarization as a function
of photon energy for disk models with various disk inclinations
$i_\disk$ and black hole spins $a_*$.  Following \citet{con80}, we
have set the black hole mass $M=9M_\odot$ and the mass accretion rate
$\dot{M}=7\times 10^{17}$ g s$^{-1}$.  The spectral hardening factor
is $f_\col=1.6$. The inner boundary of the disk is assumed to be at
the ISCO and to have a vanishing torque. The
various lines correspond to different models, as indicated.  All the
models computed by \citet{con80} have been recomputed and are shown,
along with a number of additional models.  The dotted lines correspond
to the Newtonian limit.
\label{polar_con}}
\end{figure}

\begin{figure}
\epsscale{0.9}
\plotone{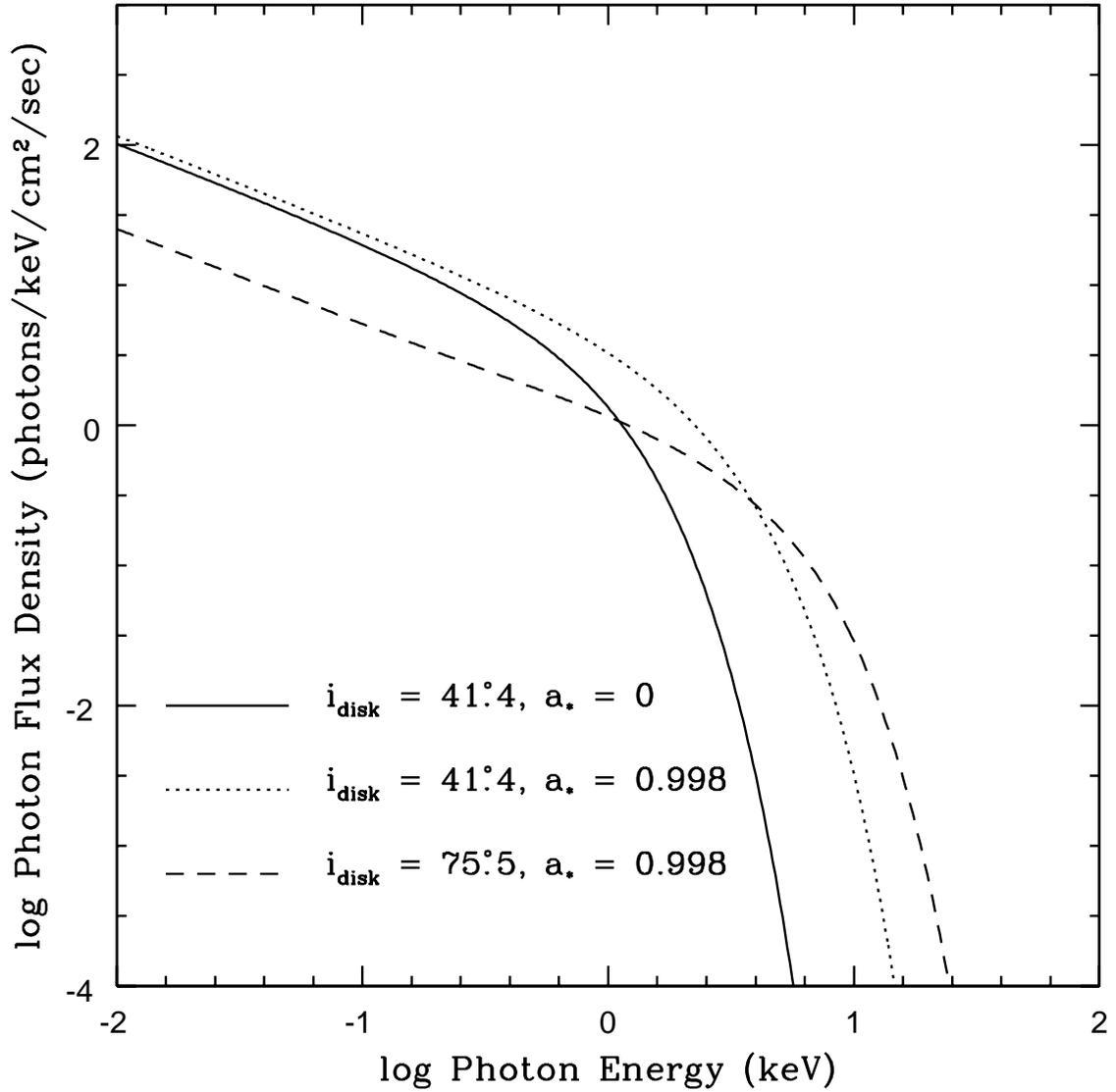}
\vspace{0.8cm}
\caption{Continuum disk spectra computed with the publicly-available
relativistic accretion disk model {\sc kerrbb} (Li et al. 2005) for
the three disk models shown in Fig.~7 of \citet{con80}.  The distance
to the source is assumed to be 10 kpc and the spectral hardening
factor is taken to be $1.6$.
\label{connors_spec}}
\end{figure}

Figure \ref{polar_con} shows the variation of the observed
polarization angle $\langle\psi\rangle$ and the observed degree of
polarization $\langle P\rangle$ as a function of photon energy for two
disk inclinations $i_\disk$ and various values of the dimensionless
spin parameter $a_*=cJ/GM^2$, where $J$ is the angular momentum of the
black hole.  All the models shown in Fig. 8 of Connors et al. (1980)
are included here; several other models are also shown for
completeness.  Qualitatively, our results are similar to those
obtained by Connors et al. (1980).  Quantitatively, however, there are
unexpectedly large differences.  Compared to Fig.~8 of \citet{con80},
our results in Fig.~\ref{polar_con} predict less photon absorption at
low photon energies (indeed this can be seen clearly in the Newtonian
limit, the dotted lines in the figure). The results for high photon
energy are also different.\footnote{Note that we have set
$\dot{M}=7\times10^{17} \,{\rm g\,s^{-1}}$ in the models shown in
Figs.~\ref{polar_con} and \ref{connors_spec}, since this is the value
mentioned by Connors et al. (1980) in the second sentence of \S~IVc.
In their caption to Fig. 5, however, they mention
$\dot{M}=10^{17}\,{\rm g\,s^{-1}}$.  We have computed models with the
latter value of $\dot{M}$ (not shown) and the results are nearly the
same.}

Rather surprisingly, the models differ considerably even in the basic
shape of the continuum spectrum, as seen by comparing Fig.~7 of
Connors et al. (1980) with the corresponding results from our model
shown in Fig.~\ref{connors_spec}.  We have assumed that the disk
emission is purely blackbody, and we have allowed for the spectral
hardening due to electron scattering and Comptonization through a
constant hardening factor $f_\col$.  The spectra shown by Connors et
al.  do not appear to be pure blackbody (see their Fig.~7).  Indeed,
their model seems to have too much emission both at low and high
energies.\footnote{At low photon energies, the blackbody photon flux
density should have the well-known asymptotic form $N_{E}\propto
E^{-2/3}$, e.g., our Fig.~\ref{connors_spec}, or equivalently the
energy spectrum should go as $F_E\propto E^{1/3}$.  However, Fig.~7 of
\citet{con80} seems to behave as $N_{E}\propto E^{-1}$ near $E=0.1$
keV.  Also, their model C, which corresponds to a rapidly spinning
black hole ($a_*=0.998$) has an unphysically hard spectrum with
emission extending up to 100 keV, which is much too hot for thermal
emission from a thin accretion disk around a stellar-mass black hole.}
Since we do not have sufficient information on the precise assumptions
made by \citet{con80} in their calculations, we are unable to resolve
the discrepancy.

We have also compared our results with the calculations reported by
\citet{ago00} and \cite{dov08}.  The easiest and most direct comparison 
is when returning radiation and photon absorption are neglected, which 
has been assumed in the model of \cite{dov08}, amd corresponds to the 
dashed lines in Fig. 11 of \citet{ago00}.  For this case, we find very 
good agreement with the results from our code for all quantities of
interest: flux, degree of polarization, and polarization angle.  When
returning radiation is included, Agol \& Krolik (2000) calculute the
polarization induced by scattering off the surface of the disk.  This
effect is currently neglected in our model.  We also neglect the
effect of Faraday rotation due to magnetic fields in the disk
atmosphere \citep{ago98}.

Before concluding this subsection, we would like to note the following
interesting point.  Figure~\ref{polar_con} shows that, for a fixed $M$
and $i_\disk$, the polarization of the observed radiation exhibits a
large variation as a function of the spin parameter.  Therefore, if we
have already measured $M$ and $i_\disk$ (see the next two
subsections), then we could use polarization data to constrain $a_*$.
This would provide an estimate of $a_*$ independent of the disk
continuum fitting method.  In fact, with sufficiently high quality
polarization data we might even be able to solve for both $i_\disk$
and $a_*$ without using the continuum spectrum at all.

\subsection{Spin-Inclination Degeneracy in the Disk Continuum Spectrum}
\label{degen}

We proceed now to discuss the continuum fitting method and the
important role that polarization measurements could play in this
method.  To determine the spin parameter $a_*$ of a black hole in an
X-ray binary, we fit the X-ray continuum spectrum of its accretion
disk using the publicly-available model {\sc kerrbb}.  Three
parameters of the system need to be measured independently: the mass
of the black hole $M$, the distance to the black hole $D$, and the
disk inclination angle $i_\disk$.  {\sc kerrbb} has two principal fit
parameters: the disk mass accretion rate $\dot M$ and the
dimensionless spin parameter $a_*$.  The mass accretion rate serves
the role of a normalization factor and depends on the X-ray
luminosity, while $a_*$ is determined principally by the temperature
or hardness of the X-ray spectrum.  Generally, it is straightforward
to estimate these two parameters from X-ray continuum data.  In fact,
even when the spectral data require additional features for a good
fit, e.g., photoelectric absorption, an iron line, a weak Comptonized
tail, one can still estimate $\dot M$ and $a_*$ with good precision.

In addition to $M$, $D$ and $i_\disk$, the spin determination requires
a theoretical model of the disk flux profile $F(R)$ as a function of
radius $R$, and a detailed disk atmosphere model to estimate the
spectral hardening factor $f_\col$.  For $F(R)$, {\sc kerrbb} makes
use of the relativistic disk model of Novikov \& Thorne (1973).  The
no-torque inner boundary condition assumed in this model has been
questioned \citep[e.g.,][]{kro99,kro02,kro05}, but recent work suggests 
that it may not be a serious issue when the disk luminosity is low and 
the disk is geometrically thin \citep{pac00,li02,afs03,sha08a,sha08b}.  
With regard to $f_\col$, disk atmosphere models
including metal opacities have been recently computed and $f_\col$ has
been estimated as a function of the disk luminosity and inclination
\citep{dav05,dav06a,dav06b}.  This is a significant improvement over
previous models which included only free-free opacity \citep{shi95}.

We thus expect that a reliable determination of $a_*$ of a stellar black
hole can be made when we have high quality X-ray data in the thermal
state --- usually not a problem --- when we have good measurements of
$M$, $D$ and $i_\disk$.  For a number of stellar black holes $M$ and $D$
are well measured \citep{cha06,oro03,oro07}.  However, the inclination 
angle of
the inner disk cannot be directly measured, except in a few cases when
the system has a radio jet and the orientation of the jet (which is
presumably perpendicular to the inner disk plane) can be determined
\citep{hje95,hje00,oro01}.  For the remaining systems, what we can
measure is the inclination of the binary orbit $i_\orb$, and we have
to assume that the inner disk is aligned with the binary orbital
plane: $i_\disk=i_\orb$.  This assumption is probably reasonable (see
\S~\ref{inclination}), but it would be better if we could avoid it
altogether.

\begin{figure}
\epsscale{0.9}
\plotone{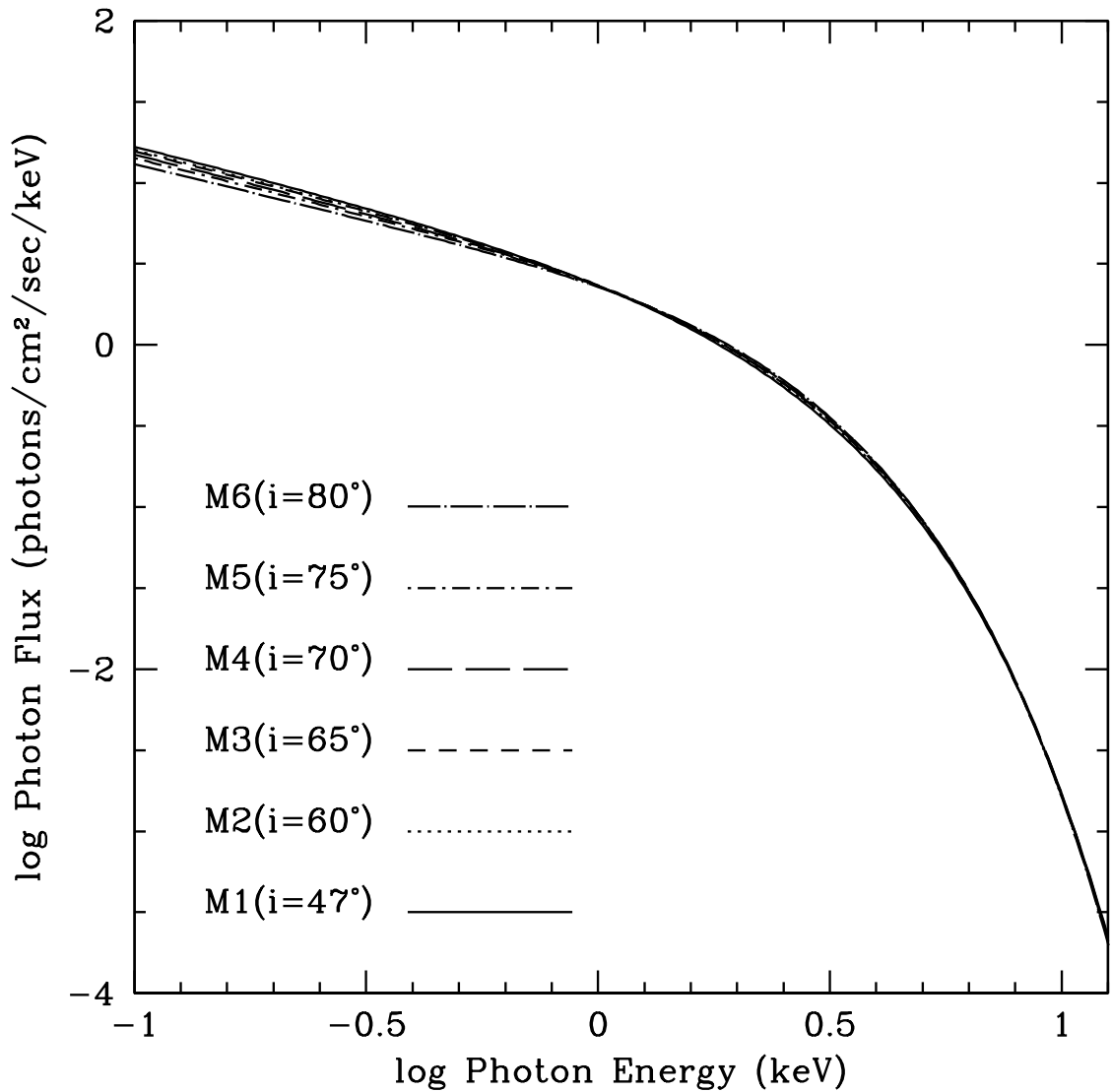}
\vspace{0.8cm}
\caption{ Disk continuum spectra computed with {\sc kerrbb} for the
models M1--M6 defined in Table~\ref{d_model} and identified by the
disk inclination angle $i$.  Although the black hole spin parameter is
very different in the various models, the continuum spectra are nearly
indinstinguishable.  This shows that the X-ray continuum spectrum
alone cannot be used to determine both the spin parameter and the disk
inclination.
\label{degeneracy}}
\end{figure}

In view of the uncertainty in the disk inclination, could we obtain
$i_\disk$ directly from X-ray spectral data?  Unfortunately, this is
not possible since a given observed continuum spectrum can be fitted
well with different combinations of $a_*$ and $i_\disk$.
Table~\ref{d_model} shows an example of a set of degenerate models
with distinct combinations of $a_*$ and $i_\disk$.  The mass of the
black hole is fixed at $10 M_\odot$, the distance to the black hole is
fixed at $10$ kpc, the spectral hardening factor is assumed to be
fixed at $1.6$, and the mass accretion rate $\dot M$ of each model has
been adjusted such that all the models have the same luminosity.
Figure~\ref{degeneracy} shows the disk continuum spectra of these
models.  Although the models have very different values of $a_*$ and
$i_\disk$, we see that their continuum spectra are nearly
indistinguishable.  Therefore, the spectrum alone is insufficient to
solve for $i_\disk$.  As we now show, polarization can very
effectively break the degeracy.

\subsection{Resolving the Spin-Inclination Degeneracy via Polarization 
Measurements}
\label{b_degen}

\begin{figure}
\epsscale{0.9}
\plotone{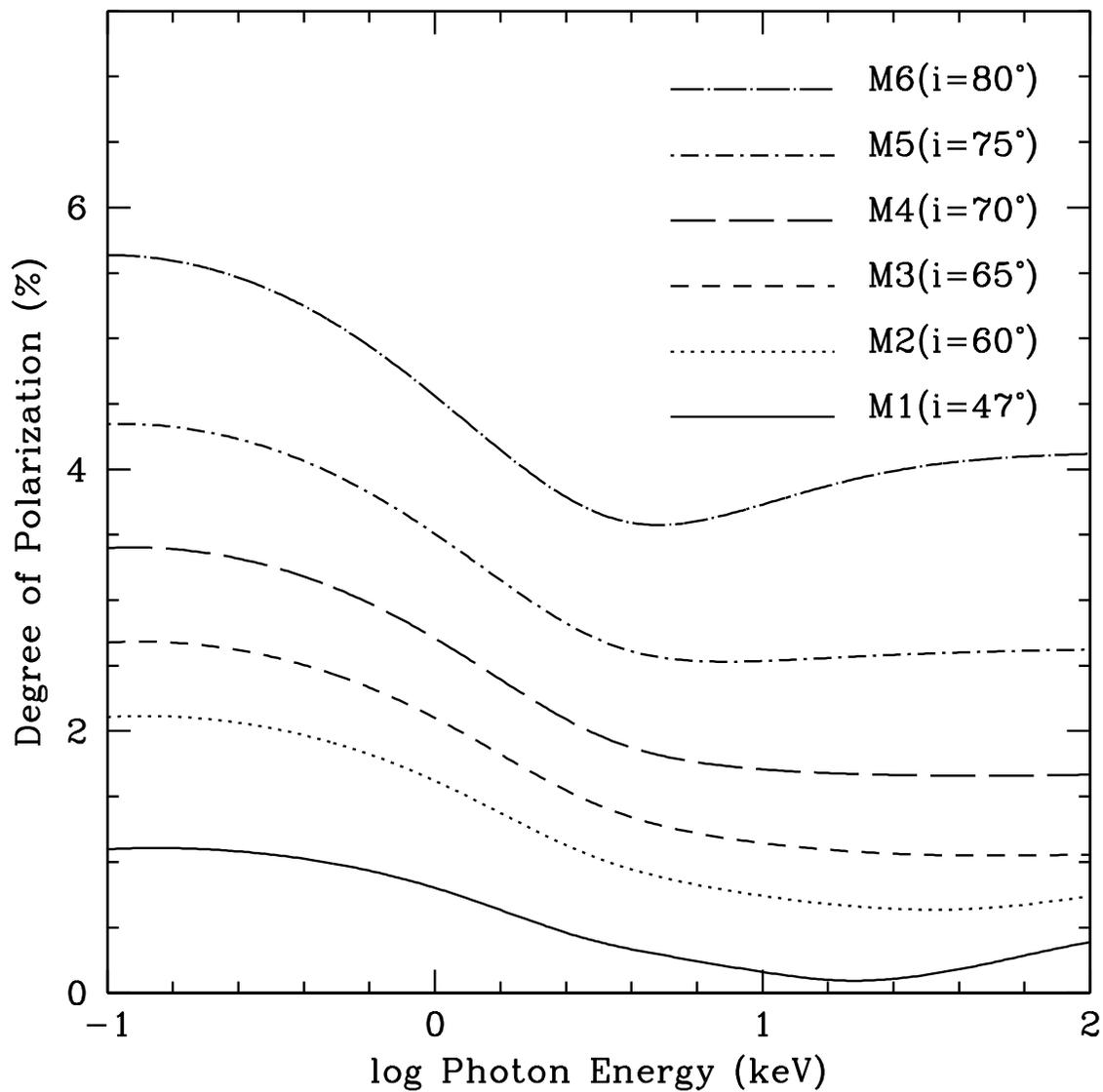}
\vspace{0.8cm}
\caption{Degree of polarization versus photon energy for the set of
degenerate disk models M1--M6 defined in Table \ref{d_model} and
identified by the disk inclination angle $i$.  The computations were
done using an extended version of {\sc kerrbb}.  In contrast to
Fig.~\ref{degeneracy}, where the continuum spectra of these models
were identical, here the models are seen to have widely different
degrees of polarization.
\label{p_degree}}
\end{figure}

Figure \ref{p_degree} shows the degree of polarization $\langle
P\rangle$ as a function of photon energy for the degenerate models
defined in Table \ref{d_model}.  We see that $\langle P\rangle$
depends sensitively on the disk inclination.  For the six models in
Table \ref{d_model}, the disk inclination angle increases
monotonically from model~M1 to model~M6, and we see that the degree of
polarization of the disk radiation also increases monotonically.  The
difference in $\langle P\rangle$ between the extreme models is quite
large.

\begin{figure}
\epsscale{0.9}
\plotone{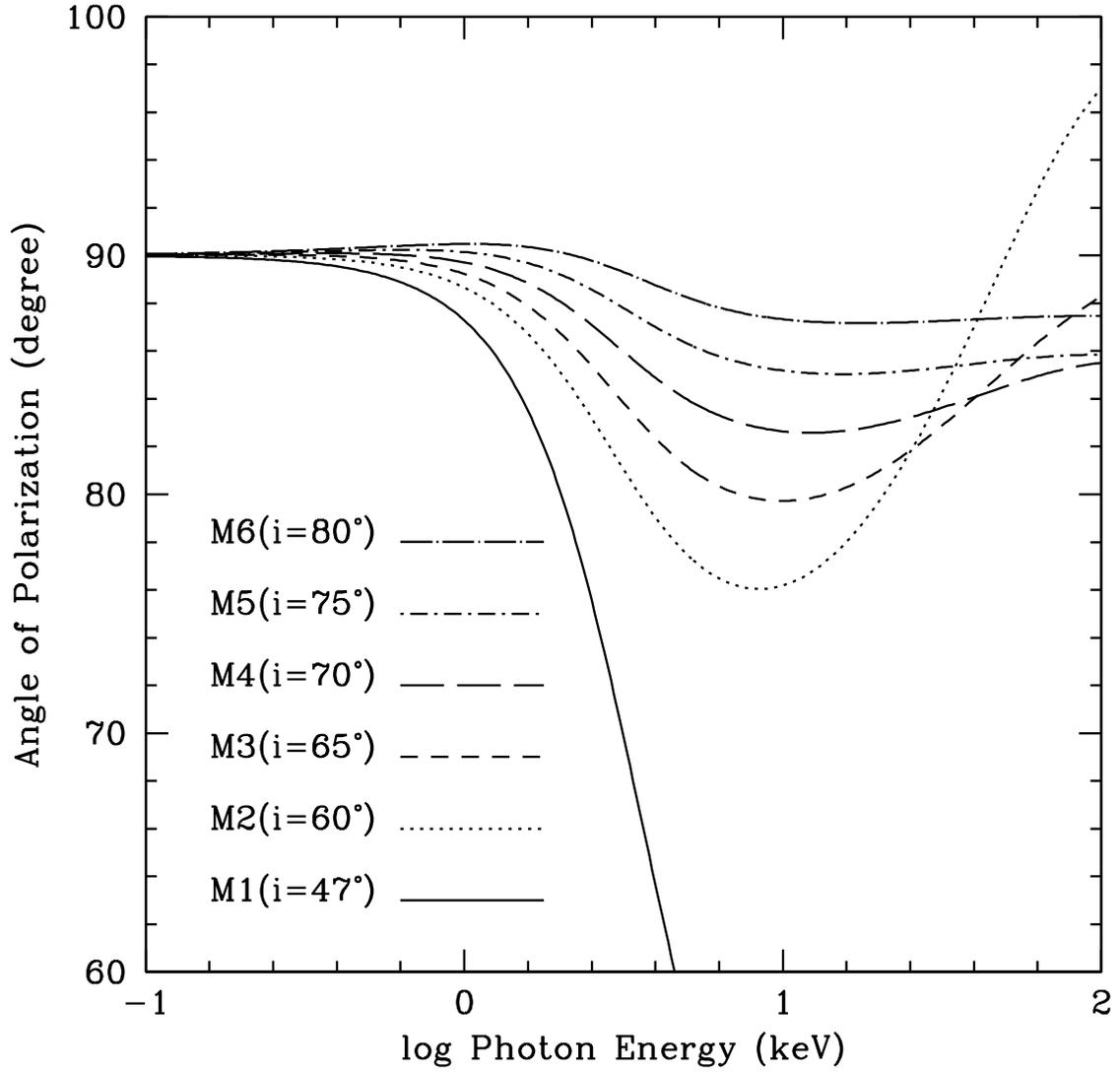}
\vspace{0.8cm}
\caption{Similar to Fig.~\ref{p_degree}, but showing the angle of
polarization.  Once again the models are easily distinguishable.
\label{p_angle}}
\end{figure}

Figure~\ref{p_angle} shows corresponding results for the polarization
angle $\langle\psi\rangle$.  Once again we see a large variation as a
function of disk inclination.  The variations are especially dramatic
at photon energies on the order of several keV.  Since the thermal
emission from the disk peaks in this region of the spectrum (and
photoelectric absorption is unlikely to be important), the effect
could be easily measured (see \S~6).

\begin{figure}
\epsscale{0.9}
\plotone{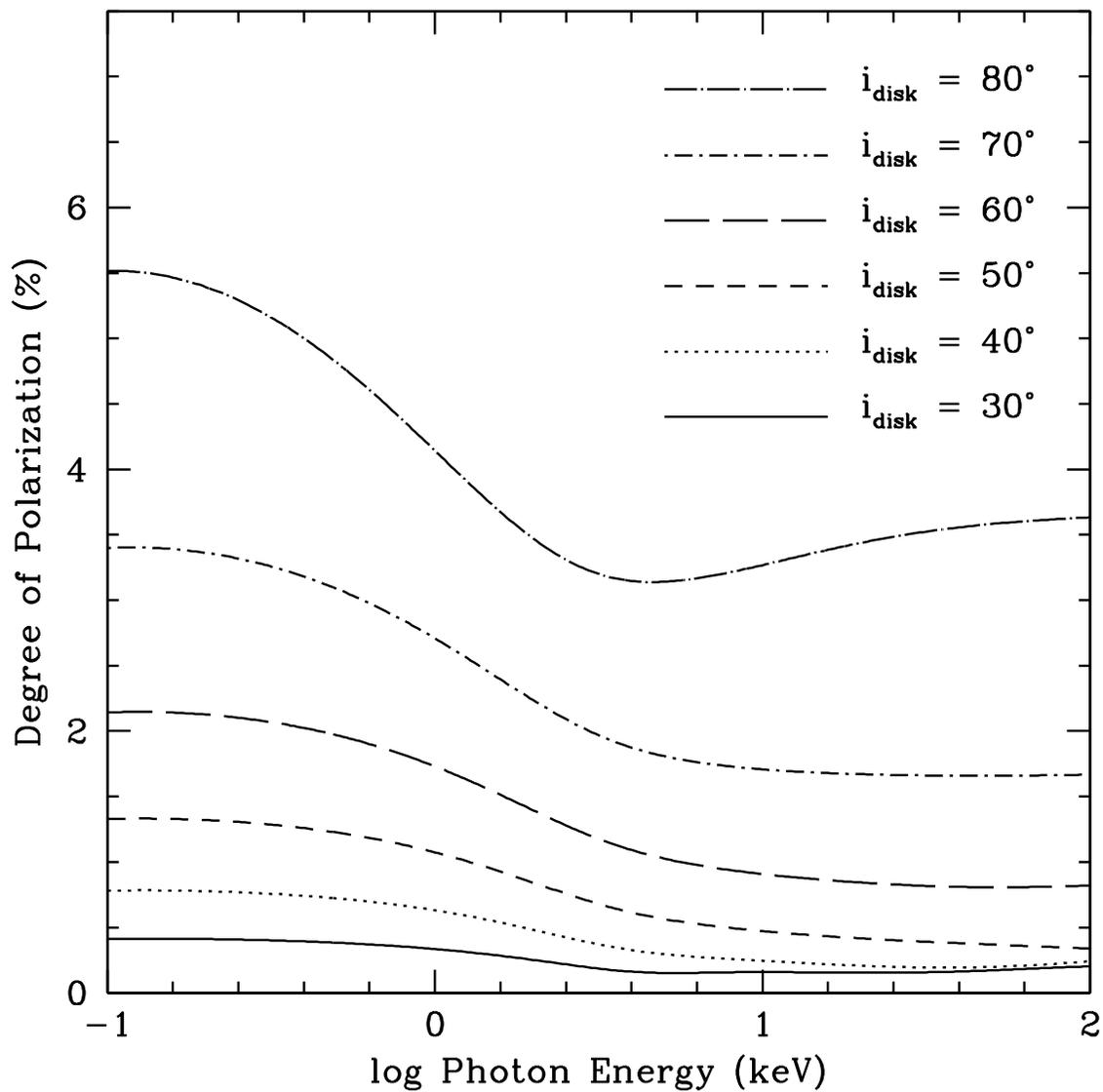}
\vspace{0.8cm}
\caption{Degree of polarization versus photon energy for a series of
accretion disk models with varying inclination angle $i_\disk$.  All
the models have $\dot{M}=2\times10^{18} ~{\rm g\,s^{-1}}$,
$M=10M_\odot$ and $a_*=0.75$.  The large variation in the degree of
polarization with $i_\disk$ means that it should be possible to
determine the disk inclination accurately with polarization data.
\label{p_degree2}}
\end{figure}

These results demonstrate that the degeneracy between the black hole
spin and the disk inclination in the continuum blackbody spectrum of
the accretion disk can be resolved by polarization measurements.  Note
that the degree of polarization, in particular, shows large variations
with disk inclination.  To illustrate this fact, we show in
Fig.~\ref{p_degree2} the run of $\langle P\rangle$ with photon energy
for an accretion disk with $\dot{M}=2\times10^{18} ~{\rm g\,s^{-1}}$
(similar to model M4 in Table~1) and a range of inclination angles
$i_\disk$ around a black hole of mass $M=10M_\odot$ and spin parameter
$a_*=0.75$.  Over the range $30^\circ \leq i_\disk \leq 80^\circ$, we
see that $\langle P\rangle$ varies by about an order of magnitude.
Thus, an observation of the polarization would provide an accurate
determination of the inclination of the inner disk.  A fit of the
continuum spectrum would then give an accurate measurement of the spin
of the black hole.

As a by-product, such work would resolve the question of whether or
not the inner disk in a black hole X-ray binary is aligned with the
orbital plane (see \S~\ref{inclination}).  Knowing the answer to this
question would provide valuable constraints on models of core collapse
and stellar black hole formation.

\section{The Prospects for an Observational Test}
\label{observation}

Is the inner accretion disk around stellar black holes warped away
from the orbital plane (\S2.2)?  If so, is the warping severe enough
to compromise the estimates of spin obtained using the
continuum-fitting method (\S\S2.1 and 5.2)?  In this section, we show
schematically that we can expect definitive answers to these question
in the near future through polarization measurements in the X-ray
band.

X-ray polarimetry is presently a hot topic, as evidenced by the 2004
X-ray Polarimetry Workshop at Stanford 
University\footnote{http://heasarc.gsfc.nasa.gov/docs/heasarc/polar/polar.html}
and the upcoming conference to be held in Rome: The Coming Age of X-ray
Polarimetry.\footnote{http://projects.iasf-roma.inaf.it/xraypol/}
Furthermore, several polarimetric missions have been proposed recently.
We now briefly consider the performance of two very different instrument
concepts developed within the severe constraints of a NASA SMEX-class
payload.  These are modest instruments because the mission costs for
a SMEX (Small Explorer) are capped at \$105 million (excluding the
launch vehicle).

The instrument considered by a team at the Goddard Space Flight Center
(GSFC) is a broadband (1--10 keV) polarimeter, which measures the
angular distribution of the tracks of the photoelectrons (Costa et al.\
2001; Swank et al.\ 2004).  The other instrument concept was developed
by a team at the Smithsonian Astrophysical Observatory (SAO).  It is a
narrowband, Bragg-crystal instrument, which operates at 2.6 keV and 5.2
keV (E. Silver 2008, private communication) and is a modern version of
the crystal polarimeter flown aboard {\it OSO-8} (Weisskopf et al.\ 1978).
The sensitivity of these two instruments is nearly identical and is
summarized in Table 2.  As indicated in the table, either instrument
can, for example, detect polarization in a 1 Crab source at the 0.3\%
level in 1 day and at the 0.1\% level in 10 days.

Our focus here is the thermal dominant (TD) state of black hole binaries
(\S2.1), or an intermediate or steep power-law (SPL) state with a strong
thermal component (McClintock \& Remillard 2006), which we here refer to
as a near-TD state.  The dominance in this state of the thermal
component makes it ideally suited for the determination of the
inclination of the inner disk via polarimetry.  Fortunately, nearly all
of the many black-hole transient sources display a thermal dominant
spectrum for weeks at a time during their outburst cycle.  Furthermore,
the sources are bright during this thermal phase, which makes them
feasible polarimetric targets.  Table 3 lists a selected sample of the
more than 40 known transient and persistent black hole systems
(McClintock \& Remillard 2006; Remillard \& McClintock 2006; Orosz et
al.\ 2007; Silverman \& Filippenko 2008).  The average intensities and
outburst durations are restricted to those times when the source was in
the TD or near-TD state.  These data are only crude characterizations of
the actual intensities/durations; see the references for precise
information.

As the data in Table 3 indicate, many sources have maintained a
TD spectrum with an intensity of $\sim 1$ Crab for
several tens of days.  Therefore, a modest instrument like those
described above would have the capability to measure the predicted
polarizations of up to $\sim 5$\% (Fig.~\ref{p_degree2}) with a
sensitivity of $\sim 0.1$\% for several transient (and persistent)
sources during the course of a mission.  Furthermore, polarization
measurements of this precision could in many cases be repeated several
times during the $\sim 50$--$100$~d thermal decay phase of a typical
transient source.

Even a single $\sim 1$--$10$ d observation would allow the inclination of
the inner disk, for a wide range of inclinations, to be determined to
roughly one degree (E. Silver 2008, private communication).  Meanwhile
the inclination angle $i_{\rm orb}$ of the orbital plane is already
known for most of the sources in Table 3 to a few degrees (Charles \&
Coe 2006; Orosz et al.\ 2003).  Thus, a simple comparison of the values
of $i_{\rm disk}$ and $i_{\rm orb}$ for a source in question will
provide a stringent test of whether its disk is warped.

\section{Discussion and Conclusions}
\label{conclusion}

In order to secure the measurement of black hole spin using the
continuum-fitting method, it is essential to obtain an independent
determination of the inclination of the inner accretion disk $i_{\rm
disk}$.  We have shown that this appears to be entirely feasible via
X-ray polarimetry.  Our models predict polarizations of up to $\sim 5$\%
that vary monotonically with inclination, while sensitivities of $\sim
0.1$\% are achievable with quite modest SMEX-class polarimeters.  Such
instruments are expected to be capable of routinely making measurements
of $i_{\rm disk}$ over a wide range of inclinations with a precision of
$\sim 1^{\circ}$.  Meanwhile, with full attention it will be possible to
achieve measurements of similar quality for the orbital inclination
angle $i_{\rm orb}$, given recent advances in optical/NIR
instrumentation (e.g., Vernet et al.\ 2007) and adaptive optics (e.g.,
van Dam et al.\ 2006).  Once reliable values of these two inclination
angles have been obtained for a sample of black hole binaries, the
question will simply be: is $i_{\rm disk} \approx i_{\rm orb}$, or is it
not?

There are of course significant hurdles that must be cleared.  For
example, the model of the disk atmosphere presented here includes only
an approximate treatment of the effects of absorption.  In the manner
of Davis \& Hubeny (2006), one must model the non-LTE effects, Compton
scattering, and the opacities due to ions of the abundant elements.
Magnetohydrodynamic models of thin disks in general relativity, which
include radiation and polarization, are the ultimate theoretical goal,
and progress is now being made on this front 
\citep{bec08,nob08,rey08,sha08b}. Then, there are observational 
complications to consider.  For
example, even in the thermal dominant state there exists some remnant
scattered coronal emission, which must be modeled.  Some other
examples of possible nettlesome sources of polarization include
self-irradiation and reflection in the disk, scattering by
interstellar grains, and the effects of a global component of magnetic
field in the disk.

Clearly, the greatest hurdle is getting an X-ray polarimeter into
space.  Thirty years has passed since the tiny {\it OSO-8} crystal
instrument successfully measured the polarization of the Crab Nebula
at 2.6 keV to be $19.2 \pm 1.0$\% at a position angle of $156\fdg4 \pm
1\fdg4$ (Weisskopf et al.\ 1978).  It is surely now time to open the
polarimetric channel for serious exploration and discovery.  In this
paper we have shown how a polarimeter can secure the measurement of
black hole spin.  This is just one exciting application of X-ray
polarimetry and there are numerous others.  

In the study of black holes, polarimetry furthermore promises to define
the geometry of the emitting elements and constrain source models in
decisive ways (\S1).  This is true, for instance, for modeling the
relativistic iron line (Miller 2007), or relativistic jets (Mirabel \&
Rodr\'iguez 1999), or the mysterious power-law component that extends
unbroken to MeV energies (Grove et al.\ 1998), or the global
oscillations that sometimes modulate up to one quarter of the total
accretion power (McClintock \& Remillard 2006).  In modeling these and
other black-hole phenomena, a central question is the geometry of the
Comptonizing coronal source, which is vaguely and variously described as
a sphere or a slab or a lamp post.  Polarimetry will provide the best,
and often only, clue to the actual geometry of the corona and other key
structures (M\'esz\'aros et al.\ 1988; Blandford et al.\ 2002).

\acknowledgements

This work was supported in part by NASA grants NNH07ZDA001N and
NNX08AJ55G and NSF grant AST-0805832.  We thank Eric Silver and 
Bill Forman for providing information on their instrument concept 
and Martin Elvis for a discussion on AGN.

\begin{appendix}

\section{Relativistic Thin Accretion Disk Model}
\label{disk_model}

We make use of a simple $\alpha$-viscosity model to describe the disk
(Shakura \& Sunyaev 1973), modified to include relativistic effects
(Novikov \& Thorne 1973; Page \& Thorne 1974).  Below, we first
discuss the relativistic terms and then proceed to describe the disk
model.

\subsection{The Disk Height}

By equation (A8) of \citet{mcc06}, the vertical gravitational
acceleration in the comoving frame of the fluid, for small excursions
$z$ from the disk mid-plane, is given by
\begin{eqnarray}
	g_z = \xi\Omega_{\kk0}^2 z \;, \label{g_z}
\end{eqnarray}
where
\begin{eqnarray}
	\Omega_{\kk0} \equiv \sqrt{\frac{GM}{R^3}}   \label{om_k0}
\end{eqnarray}
is the Newtonian angular velocity of the disk, $R$ is the cylindrical
radius, and
\begin{eqnarray}
	\xi \equiv \frac{1}{\cal C}\left(1-4 a_* x^{-3}+3 a_*^2 x^{-4}
		\right) \;, \hspace{1cm} x\equiv\sqrt{\frac{R}{R_\g}} \;,
                \hspace{1cm} R_\g\equiv\frac{GM}{c^2} \;.
\end{eqnarray}
The dimensionless function ${\cal C}$, and the functions ${\cal B}$,
${\cal D}$, and ${\cal Q}$ used later below, are defined in
\citet{nov73} and \citet{pag74}.

Hydrostatic equilibrium in the vertical direction gives
\begin{eqnarray}
	\frac{c_\ss^2}{H} \approx g_z(z=H) \;, \label{vert_equi}
\end{eqnarray}
where $H$ is the height (vertical thickness) of the disk and $c_\ss$
is the thermal sound speed of the gas at the disk mid-plane (both in
the comoving frame). From equations (\ref{g_z}) and (\ref{vert_equi}),
we obtain
\begin{eqnarray}
	H \approx \frac{c_\ss}{\Omega_\vv} \;,  \label{height}
\end{eqnarray}
where the vertical disk frequency
\begin{eqnarray}
	\Omega_\vv \equiv \Omega_{\kk0}\xi^{1/2} \;. \label{omega_v}
\end{eqnarray}

\subsection{The Flux Density and the Shear Stress of the Disk}

In this subsection we set $G=c=1$ unless either of them or both appear
explicitly in an expression. With these geometrized units, we have
$R_\g = M$ and $x=(R/M)^{1/2}$.

Page \& Thorne's functions $f$ and ${\cal Q}$ \citep[eqs. 15n and 35 
of][]{pag74} are related by
\begin{eqnarray}
	f = \frac{3}{2Mx^4}\frac{\cal Q}{{\cal B}{\cal C}^{1/2}} \;.
\end{eqnarray}
Then, by equation (11b) of \citet{pag74} and equation (D11) of \citet{li05},
we have the disk flux density \citep[c.f. eq. 5.6.14b of][]{nov73}
\begin{eqnarray}
	F &=& \frac{\dot{M}}{4\pi R}f + \frac{g_\in}{4\pi R}\left(E^\dagger
		-\Omega L^\dagger\right)_\in\left(-\frac{d\Omega}{dR}
		\right)\left(E^\dagger-\Omega L^\dagger\right)^{-2}
		\nonumber\\
		&=& \frac{3G\dot{M}M}{8\pi R^3}\frac{1}{{\cal B}{\cal C}^{1/2}}
		\left({\cal Q} + \eta\epsilon_\in \frac{{\cal C}_\in^{1/2} 
		x_\in^3}{x{\cal B}^{-1}{\cal C}^{1/2}}\right)\;,
	\label{flux}
\end{eqnarray}
where $\Omega$ is the angular velocity of the disk, $E^\dagger$ and
$L^\dagger$ are, respectively, the specific energy and the specific
angular momentum of disk particles, $g_\in\equiv \eta
\epsilon_\in\dot{M}/\Omega_\in\ge 0$ is the torque applied at the
inner boundary of the disk (at the ISCO), and $\epsilon_\in\equiv 1-
E_\in^\dagger$ is the specific gravitational binding energy at the
inner boundary (hereafter the subscript `$\in$' indicates evaluation
at the disk inner boundary).

By equation (11c) of \citet{pag74}, the vertically integrated {\em
coordinate-frame component} of the shear stress tensor of the disk is
\begin{eqnarray}
	W_\varphi^R = \frac{g}{2\pi R}
		&=& \frac{1}{2\pi R}\left[\frac{E^\dagger-\Omega L^\dagger}
		{-d\Omega/dR} \dot{M}f + \frac{\left(E^\dagger-\Omega 
		L^\dagger\right)_\in}{E^\dagger-\Omega L^\dagger}
		g_\in\right]   \nonumber\\
		&=& \frac{\dot{M}}{2\pi}\left(\frac{M}{R}\right)^{1/2}
		\left({\cal Q}+\eta\epsilon_\in\frac{{\cal C}_\in^{1/2} 
		x_\in^3}{x {\cal B}^{-1}{\cal C}^{1/2}}\right) \;,
	\label{W_fir}
\end{eqnarray}
where $\varphi$ is the azimuthal coordinate in the disk.  The vertically
integrated {\em comoving-frame component} of the shear stress, $W = 2H 
t_{\hat{\varphi}\hat{R}}$ (hats indicate coordinate components in a frame 
comoving with the disk), is related to $W_\varphi^R$ by \citep[the last 
equation in page 428 of][]{nov73}
\begin{eqnarray}
	W_\varphi^R  = R {\cal B} {\cal C}^{-1/2}{\cal D}\, W \;.
\end{eqnarray}
Hence, equation (\ref{W_fir}) agrees with equation (5.6.14a) of
\citet{nov73} in the limit of $\eta = 0$ (zero torque at the inner
boundary). Then we have
\begin{eqnarray}
	W = \frac{\dot{M}}{2\pi}\left(\frac{M}{R^3}\right)^{1/2}
		\frac{{\cal C}^{1/2}}{\cal BD}\left({\cal Q} +
		\eta\epsilon_\in\frac{{\cal C}_\in^{1/2} x_\in^3}{x {\cal 
		B}^{-1}{\cal C}^{1/2}}\right)\;.
	\label{W}
\end{eqnarray}

Following \citet{sha73} we describe the viscosity in the disk via a 
dimensionless parameter $\alpha$ and assume that $t_{\hat{\varphi}\hat{R}} 
= (3/2)\alpha P_\cc$, where $P_\cc$ is the pressure at the disk mid-plane, 
and the extra factor 3/2 is in order to be similar to the Newtonian equation. 
Then we obtain
\begin{eqnarray}
	3H\alpha P_\cc = \frac{\dot{M}}{2\pi}\left(\frac{M}{R^3}\right)^{1/2}
		\frac{{\cal C}^{1/2}}{{\cal BD}}\left({\cal Q} +
		\eta\epsilon_\in\frac{{\cal C}_\in^{1/2} x_\in^3}{x {\cal 
		B}^{-1}{\cal C}^{1/2}}\right) \;.
\end{eqnarray}
Using $2H\alpha P_\cc = 2H\alpha\rho c_\ss^2 = \alpha\Sigma c_\ss^2$ and
equation (\ref{om_k0}), where $\Sigma$ is the surface mass density of the 
disk, we then have
\begin{eqnarray}
	3\pi\frac{\alpha c_\ss^2}{\Omega_{\kk0}}\Sigma = \dot{M}
		\frac{{\cal C}^{1/2}}{{\cal BD}} \left({\cal Q} +
		\eta\epsilon_\in\frac{{\cal C}_\in^{1/2} x_\in^3}{x {\cal 
		B}^{-1}{\cal C}^{1/2}}\right)\;.
	\label{alpha_disk}
\end{eqnarray}
Note that the Newtonian limit ($x\gg 1$) of equation
(\ref{alpha_disk}) is
\begin{eqnarray}
	3\pi\frac{\alpha c_\ss^2}{\Omega_{\kk0}}\Sigma = \dot{M}
		\left[1-\left(\frac{R_\in}{R}\right)^{1/2}\right] \;.
	\label{alpha_disk0}
\end{eqnarray}

Although in this paper we only consider the case $\eta=0$, we keep
$\eta$ in the above formulae for generality.

\subsection{One Zone Disk Model}

We make the standard one-zone approximation to describe the vertical
structure of the disk \citep{fra02}.  Thus, in terms of
the mid-plane pressure $P_\cc$ and density $\rho_\cc$, the thermal
sound speed and the surface mass density are given by
\begin{eqnarray}
	c_\ss^2 &=& {P_\cc/\rho_\cc} \;, \label{cs2a}\\
	\Sigma &=& 2H\rho_\cc \;.   \label{sigma}
\end{eqnarray}

To allow for the effect of scattering and absorption opacity, we use
the following approximate formula to relate the escaping flux density
$F$ to the mid-plane temperature $T_\cc$,
\begin{eqnarray}
	F = {\sigma T_\cc^4 \over (3/4)[\tau/2+1/\sqrt{3}+1/(3\tau_\ab)]} \;,
	\label{F_esc}
\end{eqnarray}
where $\sigma$ is the Stefan-Boltzmann constant, $\tau_\ab$ is the
absorptive optical depth from the mid-plane to the surface of the
disk, and $\tau$ is the total optical depth.  This formula is taken
from \citet{pop95} and is based on the work of \citet{hub90}.
Correspondingly, we use the following equation of state of the fluid,
\begin{eqnarray}
	P_\cc = {\rho_\cc kT_\cc\over\mu m_\u} + {4\sigma T_\cc^4\over 3c} 
		{\left[\tau/2+1/\sqrt{3}\over \tau/2+1/\sqrt{3}+1/(3\tau_\ab)
		\right]} \;,
	\label{pressure}
\end{eqnarray}
where the two terms correspond to gas and radiation pressure.  Here,
$k$ is the Boltzmann constant, $m_\u$ is the atomic mass unit, and
$\mu$ is the mean molecular weight.  We assume a fully ionized gas of
solar composition, $\mu=0.603$, and we approximate $\tau$ and
$\tau_\ab$ as
\begin{eqnarray}
	\tau = \tau_\es + \tau_\ab \;,
	\hspace{1cm}
	\tau_\es = 0.346\,(\Sigma/2) \;,
	\hspace{1cm}
	\tau_\ab = 6.6\times10^{22}\rho_\cc T_\cc^{-7/2}\,(\Sigma/2) \;, 
	\label{tau}
\end{eqnarray}
where the coefficients are in cgs units.

The disk solution is easily obtained using the above equations.  At a
given radius $R$ and for an assumed value of $\Sigma$, equation
(\ref{alpha_disk}) allows us to calculate $c_\ss^2$, equation
(\ref{height}) gives $H$, equation (\ref{sigma}) gives $\rho_\cc$,
equation (\ref{cs2a}) gives $P_\cc$, equations (\ref{pressure}) and
(\ref{tau}) allow us to solve for $T_\cc$, and finally equation
(\ref{F_esc}) gives the radiation flux $F$.  Equating this estimate of
the flux to the correct value given in (\ref{flux}) gives an algebraic
equation with only one unknown, $\Sigma$, which can be solved
numerically.  Once we have $\Sigma$, we immediately obtain the values
of $\tau_\es$ and $\tau_\ab$ and this allows us to calculate the
polarization suppression factor $q_\w$ in equation
(\ref{q_parameter}).

\section{Propagation of the Polarization Vector in Kerr Spacetime}
\label{p_prop}

In this section we use geometrized units: $G=c=1$.

In terms of the Newman-Penrose orthonormal (ON) tetrad
$\left\{l^a,n^a,m^a, {\overline{m}}^a\right\}$ [where $l^a$ and $n^a$
are real vectors, $m^a$ and $\overline{m}^a$ are complex, and
$\overline{m}^a = \left(m^a\right)^*$ is the complex conjugate of
$m^a$], the Kerr metric tensor can be written as
\citep{cha83,wal84}\footnote{We adopt the signature convention of
\citet{mis73} and \citet{wal84}, which is opposite to that of
\citet{cha83}.}
\begin{eqnarray}
	g_{ab} = 2\left[-l_{(a} n_{b)} + m_{(a} \overline{m}_{b)}\right] \;,
	\label{g_ab2}
\end{eqnarray}
where the use of parentheses `()' in subscripts denotes symmetrization
of a tensor.  The ON tetrad satisfies the orthogonality conditions
\begin{eqnarray}
	{\bf l}\cdot{\bf m} = {\bf l}\cdot\overline{\bf m} = {\bf n}
		\cdot{\bf m} = {\bf n}\cdot\overline{\bf m} = 0 \;,
\end{eqnarray}
the normalization conditions
\begin{eqnarray}
	{\bf l}\cdot{\bf n} = -1 \;, \hspace{1cm} {\bf m}\cdot
		\overline{{\bf m}} = 1 \;,
\end{eqnarray}
as well as the null conditions
\begin{eqnarray}
	{\bf l}\cdot{\bf l} = {\bf n}\cdot{\bf n} = {\bf m}
		\cdot{\bf m} = \overline{\bf m}\cdot\overline{\bf m} = 0 \;,
\end{eqnarray}
where ${\bf l}\cdot{\bf m} \equiv l^a m_a$, etc.  Then, for any vector
$f^a$, we have
\begin{eqnarray}
	{\bf f}\cdot{\bf f} = -2 [({\bf f}\cdot{\bf l})({\bf f}\cdot{\bf n}) 
		- ({\bf f}\cdot{\bf m})({\bf f}\cdot\overline{\bf m})] \;.
	\label{ff}
\end{eqnarray}

Two Killing vectors and one Killing tensor in a Kerr spacetime lead to
three conserved quantities along geodesics of particles and photons: the
energy, the angular momentum about the symmetric axis of the black hole,
and the square of the `total angular momentum' \citep{car68,bar72,wal84}.
The three conserved quantities enable one to solve the timelike and null 
geodesics in a Kerr spacetime \citep{cha83,li05}.

The Kerr metric -- like any other type D (II-II) vacuum spacetime --
possesses a conformal Killing spinor \citep{wal70,wal84} which enables
one to determine the parallel propagation of {\em polarization
vectors} along null geodesics in a simple manner. This is possible
because of the {\em Walker-Penrose theorem}: If {\bf k} is a null
geodesic, affinely parametrized, and {\bf f} is a vector orthogonal to
{\bf k} and parallelly propagated along it, then, in a type-D
spacetime, the quantity
\begin{eqnarray}
	K_\wp = 2 \left[({\bf k}\cdot {\bf l})({\bf f}\cdot {\bf n})
		-({\bf k}\cdot {\bf m})({\bf f}\cdot {\bf \overline{m}})
		\right]\Psi_2^{-1/3}
	\label{k_wp}
\end{eqnarray}
is conserved along the geodesic \citep{wal70,cha83}, i.e.,
\begin{eqnarray}
	k^a\nabla_a K_\wp = 0 \;.
\end{eqnarray}
In equation (\ref{k_wp}), $\Psi_2$ is the only nonvanishing Weyl scalar in
a Kerr spacetime
\begin{eqnarray}
	\Psi_2 = -\frac{Mr}{|\rho|^6}\left(r^2-3a^2\cos^2\vartheta\right)-
		i\frac{aM\cos\vartheta}{|\rho|^6}\left(3r^2-a^2\cos^2\vartheta
		\right) = -\frac{M}{{\rho^*}^3} \;,
	\label{psi_2}
\end{eqnarray}
where we have adopted Boyer-Lindquist coordinates
$(t,r,\vartheta,\varphi)$, and
\begin{eqnarray}
	|\rho|^2 \equiv \rho\rho^* = r^2 + a^2 \cos^2\vartheta \;.
        \label{msigma}
\end{eqnarray}

Two corollaries can be derived from the Walker-Penrose theorem.

{\em Corollary 1}: $K_\wp$ is invariant under the transformation ${\bf f}
\rightarrow {\bf f}+ \alpha {\bf k}$, where $\alpha$ is an arbitrary 
function.

{\em Corollary 2}: In a Kerr spacetime, the Walker-Penrose theorem implies 
that
\begin{eqnarray}
	\left|K_\wp\right|^2 = M^{-2/3} \left[{\cal Q}+(L_z-aE_\infty)^2
		\right]({\bf f}\cdot{\bf f}) \;,
	\label{kk3}
\end{eqnarray}
where $E_\infty=-k_t$ is the conserved energy-at-infinity, $L_z=k_\varphi$ 
is the conserved angular momentum about the axis of the black hole, and 
${\cal Q}=k_\vartheta^2+L_z^2\cot^2\vartheta-a^2E_\infty^2\cos^2\vartheta$.
(When $a=0$, ${\cal Q}+L_z^2$ is the square of the total angular momentum.)

Proof of Corollary 1 is straightforward by applying the identity ${\bf k}
\cdot{\bf k} = 0$ and equation (\ref{ff}).

Proof of Corollary 2 is as follows. Following the proof of the corollary 2 
in \S60 of \citet{cha83}, we have
\begin{eqnarray}
	\left|K_\wp\right|^2 = 2\left|\Psi_2\right|^{-2/3}({\bf k}\cdot
		{\bf m})({\bf k}\cdot\overline{\bf m})({\bf f}\cdot{\bf f}) 
		\;.
	\label{kk2}
\end{eqnarray}
Then, defining
\begin{eqnarray}
	S &\equiv& \frac{k_\varphi}{\sin\vartheta} +a \sin\vartheta\, k_t
		= \frac{L_z}{\sin\vartheta}-a \sin\vartheta\, E_\infty \;,
	\label{cs} \\
	T &\equiv& k_\vartheta = \sgn (k_\vartheta) \sqrt{{\cal Q} + a^2 
		E_\infty^2 \cos^2\vartheta- L_z^2\cot^2\vartheta} \;,
	\label{ct} 
\end{eqnarray}
one finds that
\begin{eqnarray}
	({\bf k}\cdot{\bf m})({\bf k}\cdot\overline{\bf m}) = \frac{1}
		{2|\rho|^2}\left(S^2+T^2\right) \;.
	\label{km_evol}
\end{eqnarray}
It can be checked that
\begin{eqnarray}
	S^2 + T^2 = {\cal Q}+(L_z-aE_\infty)^2 = \mbox{constant} \;,  
	\label{ss_tt}
\end{eqnarray}
and
\begin{eqnarray}
	\left|\Psi_2\right|^{-2/3} = M^{-2/3}|\rho|^2 \;. \label{psi22}
\end{eqnarray}
Substituting equations (\ref{km_evol})--(\ref{psi22}) into equation
(\ref{kk2}), the identity in equation (\ref{kk3}) is proved.

Let us define
\begin{eqnarray}
	K_\wp \equiv (-M)^{-1/3} \left(K_1 + i K_2\right) \;,  \label{k1_k2}
\end{eqnarray}
where $K_1$ and $K_2$ are real constants. Then, substituting equation
(\ref{psi_2}) into equation (\ref{k_wp}), we obtain
\begin{eqnarray}
	K_1 + iK_2 = 2\rho^* \left[({\bf k}\cdot {\bf l})({\bf f}\cdot 
		{\bf n})-({\bf k}\cdot {\bf m})({\bf f}\cdot {\bf 
		\overline{m}})\right] \;.
	\label{k1k20}
\end{eqnarray}
In terms of vector components in the Boyer-Lindquist coordinate
system, we have
\begin{eqnarray}
	K_1 +i K_2 &=& \frac{1}{r+i a \cos\vartheta} \left\{\left(r^2+a^2
		\right)\left(k_r f_t-k_t f_r\right) + a \left(k_r f_\varphi
		-k_\varphi f_r\right)\right. \nonumber\\
		&&\left.+\frac{i}{\sin\vartheta}\left[k_\vartheta f_\varphi 
		- k_\varphi f_\vartheta- a\sin^2\vartheta\left(k_t 
		f_\vartheta -k_\vartheta f_t\right)
		\right]\right\} \;,
	\label{k_wp3b}
\end{eqnarray}
where we have used the fact that
\begin{eqnarray}
	k_a f^a = 0 \;. \label{k_f}
\end{eqnarray}
Equation (\ref{k_wp3b}) is very useful since $k_t$ and $k_\varphi$ are
conserved quantities, and $k_\vartheta$ can also be expressed in terms
of conserved quantities ($E_\infty$, $L_z$, and ${\cal Q}$).
Therefore, $k_r$ can be calculated using the equation $g^{ab} k_a
k_b=0$.

To solve for the propagation of photon polarization, we take ${\bf f}$ to be 
the unit {\em polarization vector}, i.e. ${\bf f} = {\bf A}/A$ where 
${\bf A}$ is the wave amplitude vector \citep{mis73}. The vector ${\bf f}$
satisfies equation (\ref{k_f}), and
\begin{eqnarray}
	k^a\nabla_a f^b = 0 \;,    \label{grad_f}
\end{eqnarray}
which states that the polarization vector is parallelly propagated
along rays.  Although for arbitrarily polarized radiation ${\bf f}$ is
complex, for a linearly polarized beam it can be chosen to be a real
vector \citep{mis73}.  Hence, we assume that ${\bf f}$ is real, with
\begin{eqnarray}
	f_a f^a = 1 \;.    \label{unit}
\end{eqnarray}
The four-vector ${\bf f}$ is only defined to within a multiple of
${\bf k}$ since ${\bf k}$ satisfies the geodesic equation. Under a
transformation ${\bf f}\rightarrow {\bf f}^\prime = {\bf f} + \alpha
{\bf k}$, it is obvious that equations (\ref{k_f}) and (\ref{unit})
are preserved. However, for ${\bf f}^\prime$ to be a solution of the
parallel propagation equation (\ref{grad_f}), $\alpha$ must satisfy
$k^a\nabla_a \alpha = 0$, i.e.  $\alpha$ must be a constant along the
null geodesic.

Because of the Walker-Penrose theorem, we do not need to solve the 
propagation equation (\ref{grad_f}) of the polarization vector explicitly.
Instead, equations (\ref{k_wp3b}) and (\ref{k_f}) can be used to determine 
the polarization vector. By Corollary 2, equation (\ref{unit}) can be derived 
from equation (\ref{k_wp3b}). Since $f^a$ is parallel propagated along the 
geodesic of $k^a$, $f_a f^a$ and $f_a k^a$ are preserved along the geodesic. 
Hence, if at a point on the light ray equations (\ref{k_f}) and (\ref{unit}) 
are satisfied, they are satisfied everywhere on the light ray. However, 
equation (\ref{k_f}) cannot be derived from the Walker-Penrose theorem so 
it is an independent equation.

Equation (\ref{k_wp3b}) is equivalent to two real equations. Then,
together with equation (\ref{k_f}), we have three equations for the
four components of $f^a$. The vector $f^a$ is determined up to an
addition of a multiple of the wave vector $k^a$ (Corollary
1). However, this degree of freedom does not affect physical
measurements, since electromagnetic waves are transverse waves and a
multiple of $k^a$ only changes the component of $f^a$ along the light
propagation direction and the local time direction \citep[Exercise
22.12]{mis73}. Hence, an uncertainty arising from a multiple of $k^a$
does not prohibit us from making physical interpretations.  In fact,
we can make use of this {\em gauge freedom} to simplify the
calculation by choosing a convenient form of $f^a$.  Therefore,
equations (\ref{k_wp3b}) and (\ref{k_f}) are sufficient for solving
for the {\em physical components} of $f^a$. After evaluation of the
integral constants $K_1$ and $K_2$, equations (\ref{k_wp3b}) and
(\ref{k_f}) can be used to determine the component of $f^a$ at any
point on the light ray.

Since we have assumed that ${\bf f}\cdot{\bf f} = 1$, by equation 
(\ref{kk3}), (\ref{ss_tt}), and (\ref{k1_k2}) we have
\begin{eqnarray}
	K_1^2+K_2^2 = S^2 +T^2 = {\cal Q}+(L_z-aE_\infty)^2 \;.
	\label{kk22}
\end{eqnarray}

\subsection{Evaluation of $K_1$ and $K_2$ on the Disk Plane}
\label{disk}

On the disk plane we have $\vartheta = \pi/2$, $|\rho|^2 = r^2$ and
$\Psi_2 = -M/r^3$. Equation (\ref{k_wp3b}) is then reduced to
\begin{eqnarray}
	K_1 +i K_2 &=& \left(r^2+a^2\right)A_1 + a A_2 +i\left(A_3 + aA_4
		\right) \;,
	\label{k_wp_disk}
\end{eqnarray}
where
\begin{eqnarray}
	A_1 &=& r^{-1} \left(k_r f_t-k_t f_r\right) \;, \\
	A_2 &=& r^{-1} \left(k_r f_\varphi-k_\varphi f_r\right) \;,\\
	A_3 &=& r^{-1} \left(k_\vartheta f_\varphi - k_\varphi f_\vartheta
		\right)  \;, \\
	A_4 &=& r^{-1} \left(k_\vartheta f_t - k_t f_\vartheta\right) \;.
\end{eqnarray}

To evaluate the functions $A_1...A_4$ in terms of the local parameters
of photons as they leave the disk, e.g., the photon energy and the
propagation direction as measured by an observer corotating with the
disk, we will make use of two local frames: the {\em locally
nonrotating frame} $\left\{e_t^a, e_r^a, e_\vartheta^a,
e_\varphi^a\right\}$ which has a zero angular momentum, and the {\em
local rest frame} $\left\{e_{(t)}^a, e_{(r)}^a, e_{(\vartheta)}^a,
e_{(\varphi)}^a\right\}$ which corotates with the disk fluid
\citep{bar72,nov73,li05}. The two frames are related by a Lorentz
transformation
\begin{eqnarray}
	e_t^a &=& \Gamma\left[e_{(t)}^a - v_\varphi e_{(\varphi)
		}^a\right] \;, \label{ett}\\
	e_\varphi^a &=& \Gamma\left[-v_\varphi e_{(t)}^a + e_{(\varphi)}^a
		\right] \;, \label{efit}\\
	e_r^a &=& e_{(r)}^a \;, \label{ert}\\
	e_\vartheta^a &=& e_{(\vartheta)}^a \label{ezt}\;,
\end{eqnarray}
where $v_\varphi$ is the azimuthal velocity of a disk particle relative 
to the locally nonrotating frame, and $\Gamma = \left(1-v_\varphi^2
\right)^{-1/2}$ is the corresponding Lorentz factor. 

\begin{figure}
\epsscale{0.9} \plotone{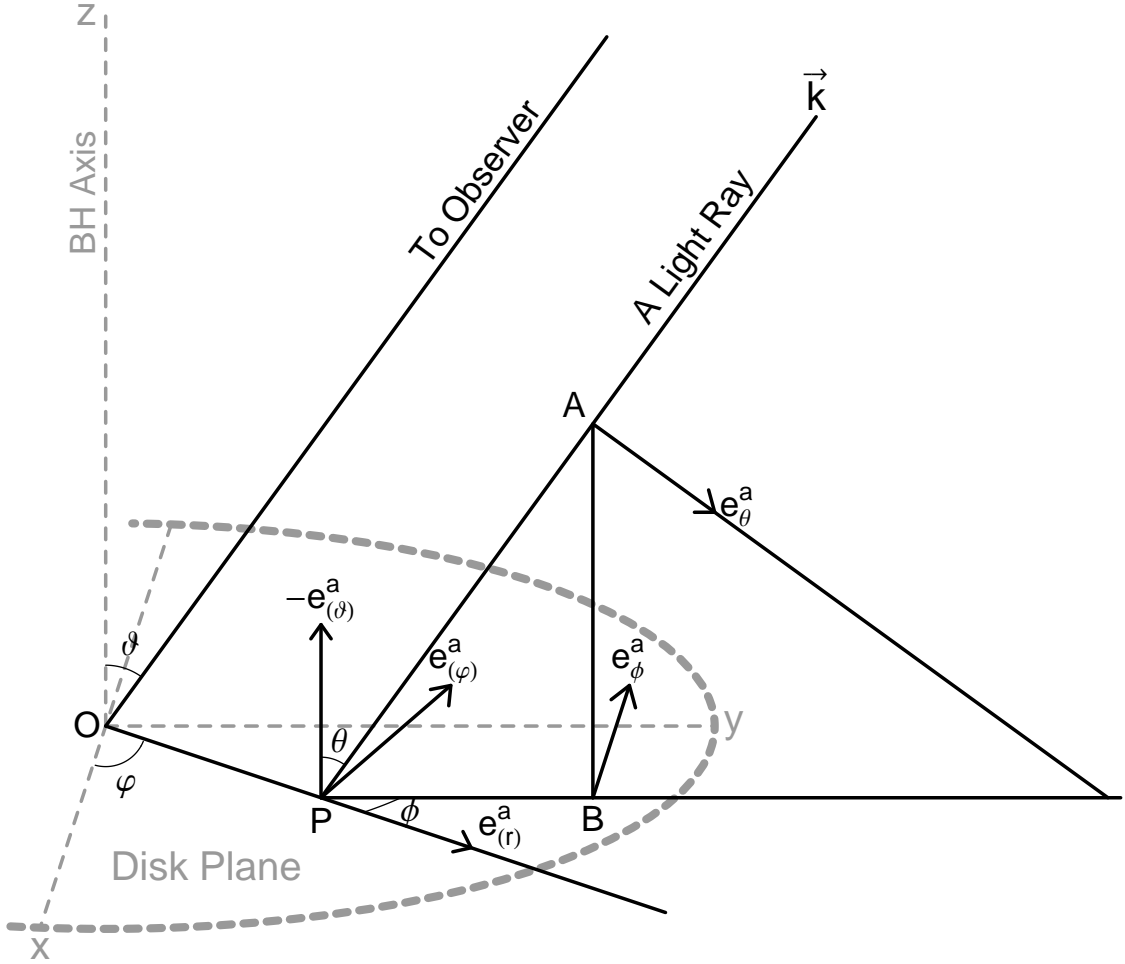}
\caption{The coordinate system on the disk plane. The disk center is located 
at $O$. $\{e_{(t)}^a, e_{(r)}^a, e_{(\vartheta)}^a, e_{(\varphi)}^a\}$ is 
a local rest frame of the disk, with $e_{(t)}^a$ not shown. A light ray is 
emitted by a disk particle at $P$, propagates along a direction labeled by 
$\vec{k}$. The plane $PAB$ is a meridian plane at $P$, perpendicular to the 
disk plane. In the spherical coordinate system $(\theta,\phi)$, $\theta$ is 
measured from the disk normal $-e_{(\vartheta)}^a$ in the meridian plane, and 
$\phi$ is measured from $e_{(r)}^a$ in the disk plane. The unit vectors 
$e_\theta^a$ (in the meridian plane and perpendicular to $\vec{k}$) and 
$e_\phi^a$ (in the disk plane and perpendicular to $PB$) are shown.
\label{coord2}}
\end{figure}

The direction of the velocity of a photon as it crosses the disk plane is
specified by the normalized four-wavevector of the photon, $n^a \equiv k^a/
k^{(t)} =k^a/E_\l$, where $k^a$ is the four-wavevector of the photon, and $E_\l
= k^{(t)} = -k_a e_{(t)}^a$ is the energy (frequency) of the photon measured
in the local rest frame. The components of $n^a$ in the local rest frame of 
the disk are
\begin{eqnarray}
	n^{(t)} = 1 \;, \hspace{0.9cm} n^{(\vartheta)} = -\cos\theta \;,
		\hspace{0.9cm} n^{(r)} = \sin\theta\cos\phi \;,
    		\hspace{0.9cm} n^{(\varphi)} = \sin\theta\sin\phi \;, 
	\label{nka}
\end{eqnarray}
where $(\theta,\phi)$ are spherical coordinates in the local rest
frame, with the polar angle $\theta$ measured from the normal to the
disk, and the azimuthal angle $\phi$ measured relative to the vector
$e_{(r)}^a$ along the disk radial direction (see
Fig. \ref{coord2}).\footnote{Note, by definition, $e_\vartheta^a$
points inward to the disk.}

We write the polarization vector at the disk surface as
\begin{eqnarray}
	f^a = f^\theta e_\theta^a + f^\phi e_\phi^a 
		&=& \left(f^\theta \cos\theta\cos\phi - f^\phi \sin\phi
		\right) e_{(r)}^a + f^\theta \sin\theta\, e_{(\vartheta)}^a
                \nonumber\\
		&&+ \left(f^\theta \cos\theta\sin\phi + f^\phi \cos\phi
		\right) e_{(\varphi)}^a \;. 
	\label{fa_disk}
\end{eqnarray}
By equation (\ref{nka}), the wave vector is given by
\begin{eqnarray}
	k^a = E_\l\left[e_{(t)}^a + \sin\theta\cos\phi\, e_{(r)}^a-\cos\theta
		\, e_{(\vartheta)}^a +\sin\theta\sin\phi\, e_{(\varphi)}^a
		\right] \;.
	\label{ka_disk}
\end{eqnarray}
It can be checked that $k_a f^a = 0$ is satisfied.
The condition $f_a f^a=1$ implies that
\begin{eqnarray}
	\left(f^\theta\right)^2 + \left(f^\phi\right)^2 = 1 \;.
	\label{fdisk_unit}
\end{eqnarray}
Although with equation (\ref{fa_disk}) the polarization vector is defined
up to an addition of a multiple of $k^a$, Corollary 1 indicates that this
does not affect the values of $K_1$ and $K_2$.

The disk angular velocity $\Omega$ is related to the linear circular velocity
$v_\varphi$ by
\begin{eqnarray}
	\Omega = \omega_0 + \chi_0\left(\frac{r^2}{A_0}\right)^{1/2} 
		v_\varphi \;,
        \label{om_disk}
\end{eqnarray}
where $A_0\equiv r^4+a^2 r(r+2M)$, $\chi_0\equiv \left(r^2\Delta/A_0
\right)^{1/2}$ (the lapse function in the equatorial plane; $\Delta\equiv 
r^2-2Mr+a^2$), and $\omega_0\equiv 2Mar/A_0$ (the frame dragging angular 
velocity in the equatorial plane). The specific angular momentum and the 
specific energy of disk particles are respectively
\begin{eqnarray}
	L^\dagger = \Gamma v_\varphi\left(\frac{A_0}{r^2}\right)^{1/2}
		\;,  \hspace{1cm}
	E^\dagger = \Gamma\chi_0 + \omega_0 L^\dagger \;.
	\label{ener_disk}
\end{eqnarray}

With the above equations and the definition of the locally nonrotating
frame \citep{bar72,li05}, we obtain
\begin{eqnarray}
	A_1 &=& \frac{E_\l}{\Delta^{1/2}}
		\left\{f^\theta\,E^\dagger\cos\theta\cos\phi-f^\phi
		\left[\Gamma\Omega\left(\frac{A_0}{r^2}\right)^{1/2}
		\sin\theta+E^\dagger\sin\phi\right]\right\} \;,\\
	A_2 &=& \frac{E_\l}{\Delta^{1/2}}
		\left\{-f^\theta\,L^\dagger\cos\theta\cos\phi
		+f^\phi\left[\Gamma\left(\frac{A_0}{r^2}\right)^{1/2}
		\sin\theta+L^\dagger\sin\phi\right]\right\} \;,\\
	A_3 &=& -E_\l
		\left\{f^\theta\,\left[L^\dagger\sin\theta +\Gamma
		\left(\frac{A_0}{r^2}\right)^{1/2}\sin\phi\right]
		+f^\phi\,\Gamma\left(\frac{A_0}{r^2}\right)^{1/2}
		\cos\theta\cos\phi\right\} \;,\\
	A_4 &=& E_\l
		\left\{f^\theta\,\left[E^\dagger\sin\theta +\Gamma\Omega
		\left(\frac{A_0}{r^2}\right)^{1/2}\sin\phi\right]
		+f^\phi\,\Gamma\Omega\left(\frac{A_0}{r^2}\right)^{1/2}
		\cos\theta\cos\phi\right\} \;.
\end{eqnarray}
Let us define
\begin{eqnarray}
	X &\equiv& \frac{1}{\Gamma r}\left(\frac{A_0}{\Delta}\right)^{1/2}
		\left[\left(r^2+a^2\right)E^\dagger - a L^\dagger\right] 
		~=~ r^2 + a^2 -a\Delta^{1/2} v_\varphi \;, \\
	Y &\equiv& \frac{A_0}{r^2\Delta^{1/2}}\left[\left(r^2+a^2\right)
		\Omega- a\right] 
		~=~ -a \Delta^{1/2} + \left(r^2+a^2\right) v_\varphi \;,
\end{eqnarray}
which satisfy
\begin{eqnarray}
	X^2 - Y^2 = \frac{A_0}{\Gamma^2} \;.    \label{x2y2}
\end{eqnarray}
Then by equation (\ref{k_wp_disk}) we have
\begin{eqnarray}
	K_1 &=& E_\l\,\frac{\Gamma r}
		{A_0^{1/2}}\left[\left(f^\theta\cos\theta\cos\phi-f^\phi
		\sin\phi\right)X - f^\phi\sin\theta\,Y\right] \;, 
	\label{k1_sol}\\
	K_2 &=& -E_\l\,\frac{\Gamma r}{A_0^{1/2}}
		\left[f^\theta\sin\theta\,Y + \left(f^\theta\sin\phi
		+f^\phi\cos\theta\cos\phi\right)X\right] \;.
	\label{k2_sol}
\end{eqnarray}
It can be checked that equation (\ref{kk22}) is satisfied.

\subsection{The Limit at Infinity}
\label{infinity}

As $r\rightarrow\infty$, we have $\chi^{-1}\approx 1+M/r$, and $\omega\approx
2Ma/r^3$. Expansion of the locally nonrotating frame at infinity leads to
\begin{eqnarray}
	e_t^a &\approx& \left(1+\frac{M}{r}\right)\left(\frac{\partial}
		{\partial t}\right)^a + {\cal O}\left(\frac{1}{r^2}
		\right) \left(\frac{\partial}{\partial t}\right)^a +
		{\cal O}\left(\frac{1}{r^2}\right) \frac{1}{r}\left(\frac{
		\partial}{\partial\varphi}\right)^a \;, \\
	e_r^a &\approx& \left(1-\frac{M}{r}\right)\left(\frac{\partial}
		{\partial r}\right)^a + {\cal O}\left(\frac{1}{r^2}
		\right) \left(\frac{\partial}{\partial r}\right)^a\;, \\
	e_\vartheta^a &\approx& \frac{1}{r}\left(\frac{\partial}{\partial 
		\vartheta}\right)^a +{\cal O}\left(\frac{1}{r^2}\right) 
		\frac{1}{r}\left(\frac{\partial}{\partial\vartheta}\right)^a\;,
		\\ 
	e_\varphi^a &\approx& \frac{1}{r\sin\vartheta} \left(\frac{\partial}
		{\partial \varphi}\right)^a + {\cal O}\left(\frac{1}{r^2}
		\right) \frac{1}{r\sin\vartheta}\left(\frac{\partial}{\partial
		\varphi}\right)^a\;.
\end{eqnarray}
Omitting all corrections at or above the order of $M^2/r^2$, the
metric takes the form
\begin{eqnarray}
	g_{ab} \approx -\left(1-\frac{2M}{r}\right) dt_a dt_b + \left(1+
		\frac{2M}{r}\right) dr_a dr_b + r^2 d\vartheta_a d\vartheta_b 
		+ r^2 \sin^2\vartheta d\varphi_a d\varphi_b \;.
\end{eqnarray}

We write the polarization vector $f^a$ in terms of the ON tetrad,
\begin{eqnarray}
	f^a = f^{\hat{r}} e_r^a + f^{\hat{\vartheta}} 
		e_\vartheta^a+ f^{\hat{\varphi}} e_\varphi^a \;,
\end{eqnarray}
where $f^{\hat{r}}$, $f^{\hat{\vartheta}}$, and $f^{\hat{\varphi}}$
must be finite. We have used the gauge freedom of $f^a$ to choose
$f^{\hat{t}} = 0$ at infinity.  Then we have, as $r\rightarrow\infty$,
\begin{eqnarray}
	f_t = 0 \;, \hspace{0.8cm}
	f_r \approx \left(1+\frac{M}{r}\right) f^{\hat{r}} \;, \hspace{0.8cm}
	f_\vartheta \approx r f^{\hat{\vartheta}} \;, \hspace{0.8cm}
	f_\varphi \approx r\sin\vartheta f^{\hat{\varphi}} \;. \label{fa_inf2}
\end{eqnarray}
Similarly, we can write the photon wave vector $k^a$ in terms of the ON 
tetrad. Expressed in $k_\mu$ (which must be finite), we have
\begin{eqnarray}
	k^a &\approx& -\left(1+\frac{M}{r}\right) k_t e_t^a + \left(1-\frac{M}
		{r}\right) k_r e_r^a + \frac{1}{r} k_\vartheta e_\vartheta^a
		+ \frac{1}{r\sin\varphi} k_\varphi e_\varphi^a 
		\nonumber\\[2mm]
		&=& - k_t e_t^a + k_r e_r^a \;,  \hspace{1cm} 
		\mbox{as $r\rightarrow\infty$} \;.
	\label{ka_inf}
\end{eqnarray}
The condition $k_a k^a = 0$ then implies that (note, $k_r>0$)
\begin{eqnarray}
	k_r = -k_t = E_\infty \;, \hspace{1cm} \mbox{as $r\rightarrow\infty$}
		\;,
	\label{kr_inf}
\end{eqnarray}
and equation (\ref{k_f}) leads to
\begin{eqnarray}
	f_r \approx -\frac{1}{r k_r}\left(k_\vartheta f^{\hat{\vartheta}}
		+ \frac{1}{\sin\vartheta} k_\varphi f^{\hat{\varphi}}\right) 
		= {\cal O}\left(\frac{1}{r}\right) \;,
	\label{fr_inf}
\end{eqnarray}
where we have omitted terms of the order $r^{-2}$.

Hence we have
\begin{eqnarray}
	\left(r^2+a^2\right)(k_r f_t-k_t f_r) &\approx& \frac{r k_t}{k_r}
		\left(k_\vartheta f^{\hat{\vartheta}}+ \frac{1}{\sin\vartheta} 
		k_\varphi f^{\hat{\varphi}}\right) + {\cal O}(1)\;, \\
	a(k_r f_\varphi-k_\varphi f_r) &\approx& a r \sin\vartheta\, k_r 
		f^{\hat{\varphi}} + {\cal O}\left(\frac{1}{r}\right) \;, \\
	k_\vartheta f_\varphi-k_\varphi f_\vartheta &\approx& r \left(\sin
		\vartheta \, k_\vartheta f^{\hat{\varphi}} - k_\varphi 
		f^{\hat{\vartheta}}\right) + {\cal O}\left(\frac{1}{r^2}
		\right) \;, \\
	a \sin^2\vartheta (k_t f_\vartheta -k_\vartheta f_t) &\approx& ar 
		\sin^2\vartheta\, k_t f^{\hat{\vartheta}} + {\cal O}\left(
		\frac{1}{r^2}\right) \;.
\end{eqnarray}
Substituting these into equation (\ref{k_wp3b}), we obtain in the
limit $r\rightarrow\infty$,

\begin{eqnarray}
	K_1 + i K_2 &=& \frac{k_t}{k_r}\left(k_\vartheta f^{\hat{\vartheta}}+ 
		\frac{1}{\sin\vartheta} k_\varphi f^{\hat{\varphi}}\right) +
		a \sin\vartheta\, k_r f^{\hat{\varphi}} \nonumber\\
		&& + i \left(k_\vartheta f^{\hat{\varphi}} - \frac{k_\varphi}
		{\sin\vartheta} f^{\hat{\vartheta}} - a \sin\vartheta\, 
		k_t f^{\hat{\vartheta}}\right) \;.
	\label{k_wp_infx}
\end{eqnarray}

Substituting $k_r = -k_t$ (eq. \ref{kr_inf}) into equation
(\ref{k_wp_infx}), we have
\begin{eqnarray}
	K_1 + i K_2 &=& -\left(T f^{\hat{\vartheta}} + S f^{\hat{\varphi}}
		\right)- i \left(S f^{\hat{\vartheta}} - T f^{\hat{\varphi}}
		\right) \;,
	\label{k_wp_inf}
\end{eqnarray}
where $S$ and $T$ are defined by equations (\ref{cs}) and (\ref{ct}), 
respectively. From equation (\ref{k_wp_inf}) we can solve for 
$f^{\hat{\vartheta}}$ and $f^{\hat{\varphi}}$,
\begin{eqnarray}
	f^{\hat{\vartheta}} = -\frac{K_2 S + K_1 T}{S^2+T^2} \;,\hspace{1cm}
	f^{\hat{\varphi}} = -\frac{K_1 S - K_2 T}{S^2+T^2} \;, \label{f_inf}
\end{eqnarray}
where $S^2 + T^2$ is given by equation (\ref{ss_tt}) and is a
constant.  From equations (\ref{kk22}) and (\ref{f_inf}), we then have
\begin{eqnarray}
	\left(f^{\hat{\vartheta}}\right)^2 + \left(f^{\hat{\varphi}}\right)^2 
		= \frac{K_1^2 + K_2^2}{S^2 + T^2} = 1 \;.
\end{eqnarray}
Since as $r\rightarrow\infty$ we have $f^{\hat{t}} = 0$ and  $f^{\hat{r}} = 0$
(eq. \ref{fr_inf}), equation (\ref{unit}) is satisfied.

\subsection{Solution of the Polarization Vector at Infinity}
\label{sol_inf}

By equations (\ref{k1_sol}), (\ref{k2_sol}), and (\ref{f_inf}) we have
\begin{eqnarray}
	f_\infty^{\hat{\vartheta}} &=& \frac{E_\infty\Gamma r}{gA_0^{1/2}(S^2
		+T^2)}\left\{f^\theta\left[S(X\sin\phi+Y\sin\theta)
		-TX\cos\theta\cos\phi\right]\right. \nonumber\\
		&&\left.+f^\phi\left[ T(X\sin\phi+Y\sin\theta)+SX\cos\theta
		\cos\phi\right]\right\} \;, \label{ftheta_inf}\\
	f_\infty^{\hat{\varphi}} &=& \frac{E_\infty\Gamma r}{gA_0^{1/2}(S^2
		+T^2)}\left\{f^\theta\left[-T(X\sin\phi+Y\sin\theta)
		-SX\cos\theta\cos\phi\right]\right. \nonumber\\
		&&\left.+f^\phi\left[ S(X\sin\phi+Y\sin\theta)-TX\cos\theta
		\cos\phi\right]\right\} \;, \label{fphi_inf}
\end{eqnarray}
where $g\equiv E_\infty/E_\l$ is the photon redshift factor. Note that, on 
the right hand sides of the above equations, $S$ and $T$ are evaluated at
the remote observer (hence $\vartheta = i_\disk$, the disk inclination
angle), but all other quantites are evaluated at the disk surface.

The solutions given in equations (\ref{ftheta_inf}) and
(\ref{fphi_inf}) can be written in a matrix form
\begin{eqnarray}
	\left(\begin{array}{c}
		f_\infty^{\hat{\vartheta}}\\[2mm]
		f_\infty^{\hat{\varphi}}
		\end{array}\right) = \left(\begin{array}{cc}
		\cos\Phi_\gr & \sin\Phi_\gr \\[2mm]
		-\sin\Phi_\gr & \cos\Phi_\gr
		\end{array}\right)\left(\begin{array}{c}
		f^\theta\\[2mm]
		f_\phi
		\end{array}\right) \;,
	\label{f_inf_matrix}
\end{eqnarray}
where
\begin{eqnarray}
	\cos\Phi_\gr ~=~ \xi_x &\equiv& \frac{E_\infty\Gamma r}{gA_0^{1/2}(S^2
		+T^2)}\left[S(X\sin\phi+Y\sin\theta)-TX\cos\theta\cos
		\phi\right] \;, 
	\label{psi_gr1} \\[1mm]
	\sin\Phi_\gr ~=~ \xi_y &\equiv& \frac{E_\infty\Gamma r}{gA_0^{1/2}(S^2
		+T^2)}\left[ T(X\sin\phi+Y\sin\theta)+SX\cos\theta
		\cos\phi\right] \;.
	\label{psi_gr2}
\end{eqnarray}
It can be checked that $\cos^2\Phi_\gr+\sin^2\Phi_\gr = 1$ is fulfilled.

Define
\begin{eqnarray}
	f_\infty^{\hat{\vartheta}} = \cos\psi_\infty \;,\hspace{1cm}
	f_\infty^{\hat{\varphi}} = \sin\psi_\infty \;,
\end{eqnarray}
and
\begin{eqnarray}
	f^\theta = \cos\psi_\emm \;,\hspace{1cm}
	f_\phi = \sin\psi_\emm \;.
\end{eqnarray}
Then equation (\ref{f_inf_matrix}) leads to
\begin{eqnarray}
	\cos\psi_\infty = \cos\left(\psi_\emm-\Phi_\gr\right) \;, \hspace{1cm}
	\sin\psi_\infty = \sin\left(\psi_\emm-\Phi_\gr\right) \;,
\end{eqnarray}
whose solution is
\begin{eqnarray}
	\psi_\infty = \psi_\emm-\Phi_\gr + 2n\pi \;, \label{psi_inf_sol}
\end{eqnarray}
where $n=0,\pm 1, \pm 2,...$.  Hence, the $\Phi_\gr$ defined by
equations (\ref{psi_gr1}) and (\ref{psi_gr2}) represents the rotation
of the polarization vector induced by the Kerr geometry and disk
rotation.

We define $\Phi_\gr$ to be the {\em primitive rotation angle}, which 
satisfies $0\le\Phi_\gr<2\pi$. Then by equations (\ref{psi_gr1}) and 
(\ref{psi_gr2}) we have
\begin{eqnarray}
	\Phi_\gr = \left\{\begin{array}{ll}
		\arccos\xi_x \;, & \mbox{if $\xi_y \ge 0$} \\
		2\pi - \arccos\xi_x \;, & \mbox{if $\xi_y < 0$} 
		\end{array}\right. \;.
	\label{psi_gr_sol}
\end{eqnarray}
We can check the nonrelativistic limit of this result in flat
spacetime. For this purpose, let us take $a=0$, $\Omega= 0$,
$v_\varphi = 0$, $\Gamma = 1$, $L^\dagger = 0$, $E^\dagger =1$,
$\chi=1$, $g=1$, and $A_0 = r^4$. Then we have $X=r^2$ and
$Y=0$. Since the light ray is not bent, we have $\theta =
\vartheta_\obs = i_\disk$.  So, we have $L_z = E_\infty
r\sin\theta\sin\phi$, $S=E_\infty r\sin\phi$, and
$T=\sgn\left(k_{\vartheta_\obs}\right)\sqrt{{\cal
Q}-L_z^2\cot^2\theta} = -E_\infty r\cos\theta\cos\phi$. Then
$\xi_x=1$, $\xi_y=0$, and hence the polarization vector is not
rotated.

The calculation of $\Psi_\gr$ and $\psi_\infty$ is simplified by using
dimensionless variables. We define dimensionless quantities by symbols with 
a tilde:
\begin{eqnarray}
	X \equiv \tilde{X} r^2 \;, \hspace{0.6cm} Y \equiv \tilde{Y} r^2 \;, 
		\hspace{0.6cm} S \equiv \tilde{S} r E_\infty \;, 
		\hspace{0.6cm} T \equiv \tilde{T} r E_\infty \;, 
		\hspace{0.6cm} A_0 \equiv \tilde{A} r^4 \;.
\end{eqnarray}
Then, by equations (\ref{psi_gr1}) and (\ref{psi_gr2}), we have
\begin{eqnarray}
	\xi_x &=& \frac{\Gamma}{g\tilde{A}^{1/2}
		\left(\tilde{S}^2+\tilde{T}^2\right)}
		\left[\tilde{S}\left(\tilde{X}\sin\phi+\tilde{Y}\sin\theta
		\right)-\tilde{T}\tilde{X}\cos\theta\cos\phi\right] \;, 
	\label{xi_x}\\
	\xi_y &=& \frac{\Gamma}{g\tilde{A}^{1/2}
		\left(\tilde{S}^2+\tilde{T}^2\right)}
		\left[\tilde{T}\left(\tilde{X}\sin\phi+
		\tilde{Y}\sin\theta\right)+\tilde{S}\tilde{X}\cos\theta
		\cos\phi\right] \;, 
	\label{xi_y}
\end{eqnarray}
where
\begin{eqnarray}
	\tilde{X} = 1 + \frac{a^2}{r^2} -\frac{a}{r}\left(\frac{\Delta}
		{r^2}\right)^{1/2} v_\varphi \;, \hspace{1cm}
        \label{X_til}
	\tilde{Y} = \left(1 + \frac{a^2}{r^2}\right) v_\varphi -\frac{a}{r}
		\left(\frac{\Delta}{r^2}\right)^{1/2} \;,
        \label{Y_til}
\end{eqnarray}
and
\begin{eqnarray}
	\tilde{S} &=& \frac{\lambda}{r\sin\vartheta_\obs}-\frac{a}{r}
		\sin\vartheta_\obs \;,
	\label{cs2} \\
	\tilde{T} &=& \sgn (k_{\vartheta_\obs})\frac{1}{r}\sqrt{Q + 
		a^2 \cos^2\vartheta_\obs- \lambda^2\cot^2
		\vartheta_\obs} \;,
	\label{ct2}
\end{eqnarray}
where $\lambda\equiv L_z/E_\infty$, and $Q\equiv {\cal Q}/E_\infty^2$ are
independent of $E_\infty$ \citep{li05}.

\end{appendix}


\clearpage
\begin{deluxetable}{cccc}
\tablewidth{0pt}
\tablecaption{Parameters corresponding to the models shown in 
Figs.~\ref{degeneracy}--\ref{p_angle}\label{d_model}}
\tablehead{
\colhead{~\hspace{0.5cm}Model\hspace{0.5cm}~} & \colhead{~\hspace{0.5cm}$a_*$\hspace{0.5cm}~} & \colhead{~\hspace{0.5cm}$i_\disk$\hspace{0.5cm}~} & \colhead{~\hspace{0.5cm}$\dot{M}$\hspace{0.5cm}~} 
}
\startdata
M1 & $0.998$ & $47.0$ & $0.48$\\
M2 & $0.900$ & $60.0$ & $1.00$\\
M3 & $0.830$ & $65.0$ & $1.40$\\
M4 & $0.750$ & $70.0$ & $2.00$\\
M5 & $0.630$ & $75.0$ & $3.20$\\
M6 & $0.450$ & $80.0$ & $5.80$\\
\enddata

\tablecomments{Explanation of symbols: $a_*$, dimensionless spin
parameter of the black hole; $i_\disk$, inclination angle of the disk
in degrees; $\dot{M}$, mass accretion rate in units of $10^{18}$ g
s$^{-1}$. Other parameters: mass of the black hole is $10 M_\odot$;
distance to the black hole is $10$~kpc; spectral hardening factor is
$1.6$. The inner boundary of the disk is at the ISCO and is assumed to 
have a zero torque boundary condition.}
\end{deluxetable}

\clearpage
\begin{deluxetable}{ccccc}
\tablewidth{0pt}
\tablecaption{Minimum Detectable Polarization (\%)\tablenotemark{a}
\label{detectability}}
\tablehead{
\colhead{~\hspace{0.3cm}Intensity (Crab)\hspace{0.3cm}~} & 
\colhead{~\hspace{0.3cm}0.001\hspace{0.3cm}~} & 
\colhead{~\hspace{0.3cm}0.01\hspace{0.3cm}~} &
\colhead{~\hspace{0.3cm}0.1\hspace{0.3cm}~} & 
\colhead{~\hspace{0.3cm}1\hspace{0.3cm}~}
}
\startdata
T$_{\rm obs} = 1$  d& 10& 3& 1&   0.3 \\ 
T$_{\rm obs} = 10$ d& 3&  1& 0.3& 0.1 \\ 
\enddata
\tablenotetext{a}{Nominal sensitivities accurate to $\approx 20$\% 
(Swank et al.\ 2004; E.\ Silver 2008, private communication).}
\end{deluxetable}

\begin{deluxetable}{llllccl}
\tablecaption{Selected Black Hole Binaries
\label{bh_binaries}}
\tabletypesize{\scriptsize}
\tablewidth{0pt}
\tablehead{
&
\colhead{Coordinate} &
\colhead{Common} &
\colhead{Year/} &
\colhead{Intensity\tablenotemark{b}} &
\colhead{Outburst\tablenotemark{d}} &
\colhead{References\tablenotemark{e}}\\
&
\colhead{Name} &
\colhead{Name/Prefix\tablenotemark{a}} &
\colhead{No.\ of Outbursts} &
\colhead{(Crab\tablenotemark{c})} &
\colhead{Duration (d)} &
}
\startdata
1&  0620--003   &(A)              &1975/1                &15               &100            &Matilsky et al.\ 1976 \\
2&  1124--684   &Nova Mus 91      &1991/1                &1                &100            &Ebisawa et al.\ 1994  \\
3&  1543--475   &(4U)             &1971/4                &1                &100            &RM06             \\
4&  1550--564   &(XTE~J)          &1998/5                &2                &100            &RM06             \\
5&  1650--500   &(XTE~J)          &2001/1                &0.5              &100            &MR06             \\
6&  1655--40    &(GRO~J)          &1994/3                &1                &300            &RM06             \\
7&  1659--487   &GX~339--4        &1972/12               &0.5              &200            &RM06             \\
8&  1705--250   &Nova Oph 77      &1977/1                &1                &50             &Watson et al.\ 1978  \\
9&  1859+226    &(XTE~J)          &1999/1                &0.5              &100            &RM06             \\
11& 2000+251    &(GS)             &1988/1                &3                &100            &Tsunemi et al.\ 1989 \\
12& 1630--472   &(4U)             &1971/16               &0.2              &100            &MR06             \\
13& 1743--322   &(H)              &1977/4                &0.5              &100            &RM06             \\
\hline
14& 0538--641   &LMC~X--3         &Persistent            &0.03             &\nodata        &MR06             \\
15& 0540--697   &LMC~X--1         &Persistent            &0.02             &\nodata        &MR06             \\
10& 1915+105    &(GRS)            &Quasi-persistent\tablenotemark{f}  &1
&\nodata        &McClintock et al. 2006 \\
16& 1956+350    &Cyg~X--1         &Persistent\tablenotemark{g}        &1                &\nodata        &Wilms et al.\ 2006      \\
\enddata
\tablenotetext{a}{A prefix to a coordinate name is enclosed in parentheses.}
\tablenotetext{b}{Crude estimate of average intensity in TD or near-TD state; see references.}
\tablenotetext{c}{1~Crab~$\approx~2.8~\times~10^{-8}$~erg~cm$^{-2}$~s$^{-1}$ (1-10 keV) for a Crab--like spectrum with $\Gamma~=~2.1$.}
\tablenotetext{d}{Crude estimate of time in TD or near-TD state; see references.}
\tablenotetext{e}{RM06 = Remillard \& McClintock 2006; MR06 = McClintock \& Remillard 2006.}
\tablenotetext{f}{In outburst since discovery in 1992; occasionally observed in TD state (McClintock et al.\ 2006).}
\tablenotetext{g}{Observed in a soft, near-TD state for extended periods (e.g., Wilms et al.\ 2006).}
\end{deluxetable}

\clearpage

\clearpage

\clearpage

\clearpage

\clearpage

\clearpage

\clearpage

\clearpage


\begin{thebibliography}{}

\bibitem[Alf\'ven \& Herlofson(1950)]{alf50} Alf\'ven, H., \& Herlofson,
        N. 1950, Phys.\ Rev., 78, 616

\bibitem[Afshordi \& Paczy\'nski(2003)]{afs03} Afshordi, N., \& Paczy\'nski, 
	B. 2003, ApJ, 592, 354

\bibitem[Agol et al.(1998)]{ago98} Agol, E., Blaes, O., \&
        Ionescu-Zanetti, C.  1998, MNRAS, 293, 1

\bibitem[Agol \& Krolik(2000)]{ago00} Agol, E., Krolik, J. H. 2000, ApJ, 
	528, 161

\bibitem[Antonucci \& Miller(1985)]{ant85} Antonucci, R. R. J., \&
        Miller, J. S. 1985, ApJ, 297, 621

\bibitem[Bardeen \& Petterson(1975)]{bar75} Bardeen, J. M., \& Petterson, 
	J. A. 1975, ApJ, 195, L65

\bibitem[Bardeen et al.(1972)]{bar72} Bardeen, J. M., Press, W. H., \&
        Teukolsky, S. A. 1972, ApJ, 178, 347

\bibitem[Beckwith et al.(2008)]{bec08} 	Beckwith, K., Hawley, 
        J. F., \& Krolik, J. H. 2008, ApJ, 678, 1180

\bibitem[Blandford et al.(2002)]{bla02} Blandford, R., Agol, E., 
        Broderick, A., Heyl, J., Koopmans, L., \& Lee, H.-W. 2002, in
	Astrophysical Spectropolarimetry, Proceedings of the XII Canary
	Islands Winter School of Astrophysics, eds. J. Trujillo-Bueno,
	F. Moreno-Insertis, \& F. S\'anchez (Cambridge: Cambridge
	Univ. Press), p. 177

\bibitem[Brenneman \& Reynolds(2006)]{bre06} Brenneman, L. W., \&
        Reynolds, C. S. 2006, ApJ, 652, 1028

\bibitem[Carter(1968)]{car68} Carter, B. 1968, Phys. Rev., 174, 1559

\bibitem[Chandrasekhar(1960)]{cha60} Chandrasekhar, S. 1960, Radiative 
	Transfer (New York: Dover)

\bibitem[Chandrasekhar(1983)]{cha83} Chandrasekhar, S. 1983, The Mathematical
	Theory of Black Holes (New York: Oxford Univ. Press) 

\bibitem[Charles \& Coe(2006)]{cha06} Charles, P. A., \& Coe,
        M. J. 2006, in Compact Stellar X-ray Sources, eds.  W. H. G. Lewin \&
        M. van der Klis (Cambridge: Cambridge Univ. Press), p. 215

\bibitem[Chen \& Eardley(1991)]{che91} Chen, K., \& Eardley, D. M. 1991, 
	ApJ, 382, 125

\bibitem[Connors \& Stark(1977)]{con77} Connors, P. A., \& Stark, R. F. 
	1977, Nature, 269, 128

\bibitem[Connors et al.(1980)]{con80} Connors, P. A., Piran, T., \& Stark, 
	R. F. 1980, ApJ, 235, 224

\bibitem[Costa et al.(2001)]{cos01} Costa, E., Soffitta, P., Bellazzini,
        R., Brez, A., Lumb, N., \& Spandre, G. 2001, Nature, 411, 662

\bibitem[Davis \& Hubeny(2006)]{dav06a} Davis, S. W., \& Hubeny, I. 2006, ApJS, 
	164, 530

\bibitem[Davis et al.(2006)]{dav06b} Davis, S. W., Done, C., \& Blaes, O. M.
	2006, ApJ, 647, 525

\bibitem[Davis et al.(2005)]{dav05} Davis, S. W., Blaes, O. M., Hubeny, I., 
	\& Turner, N. J. 2005, ApJ, 621, 372

\bibitem[Dov\v{c}iak et al.(2004)]{dov04}
	Dov\v{c}iak, M., Karas, V., Matt, G. 2004, MNRAS, 355, 1005

\bibitem[Dov\v{c}iak et al.(2008)]{dov08}
	Dov\v{c}iak, M., Muleri, F., Goosmann, R. W., Karas, V., Matt, G. 2008, 
        MNRAS, in press (arXiv:0809.0418)

\bibitem[Ebisawa et al.(1994)]{ebi94} Ebisawa, K., et al. 1994, PASJ,
        46, 375

\bibitem[Fabian et al.(2000)]{fab00} Fabian, A. C., Iwasawa, K.,
        Reynolds, C. S., \& Young, A. J. 2000, PASP, 112, 1145

\bibitem[Fragile \& Anninos(2005)]{fra05} Fragile, P. C., \& Anninos,
        P. 2005, ApJ, 623, 347

\bibitem[Fragile et al.(2007)]{fra07} Fragile, P. C., Blaes, O. M., Anninos, 
	P., \& Salmonson, J. D. 2007, ApJ, 668, 417

\bibitem[Frank et al.(2002)]{fra02} Frank, J., King, A., \& Raine, D. 2002,
	Accretion Power in Astrophysics (Cambridge: Cambridge
	Univ. Press)

\bibitem[Fryer \& Kalogera(2001)]{fry01} Fryer, C. L., \& Kalogera, V.
        2001, ApJ, 554, 548

\bibitem[Fryer \& Young(2007)]{fry07} Fryer, C. L., \& Young,
        P. A. 2007, ApJ, 659, 1438

\bibitem[Gierli\'nski et al.(1999)]{gie99} Gierli\'nski, M., Zdziarski,
        A. A., Poutanen, J., Coppi, Paolo S., Ebisawa, K., \& Johnson,
        W. N. 1999, MNRAS, 309, 496

\bibitem[Grove et al.(1998)]{gro98} 	
        Grove, J. E., Johnson, W. N., Kroeger, R. A., McNaron-Brown, K.,
	Skibo, J. G., \& Phlips, B. F. 1998, ApJ, 500, 899

\bibitem[Hjellming \& Rupen(1995)]{hje95} Hjellming, R. M., \& Rupen, M. P. 
	1995, Nature, 375, 464

\bibitem[Hjellming et al.(2000)]{hje00} Hjellming, R. M., et al. 2000, ApJ, 
	544, 977

\bibitem[Hubeny(1990)]{hub90} Hubeny, I. 1990, ApJ, 351, 632

\bibitem[King et al.(2005)]{kin05} King, A. R., Lubow, S. H., Ogilvie,
        G. I., \& Pringle, J. E. 2005, MNRAS, 363, 49

\bibitem[Krolik(1999)]{kro99}	
	Krolik, J. H. 1999, ApJ, 515, L73

\bibitem[Krolik \& Hawley(2002)]{kro02}
	Krolik, J. H., \& Hawley, J. F. 2002, ApJ, 573, 754

\bibitem[Krolik et al.(2005)]{kro05}
	Krolik, J. H., Hawley, J. F., \& Hirose, S. 2005, ApJ, 622, 1008

\bibitem[Laor et al.(1990)]{lao90} Laor, A., Netzer, H., \& Piran, T. 
	1990, MNRAS, 242, 560

\bibitem[Li(2002)]{li02} Li, L.-X. 2002, Phys. Rev. D, 67, 044007

\bibitem[Li et al.(2005)]{li05} Li, L.-X., Zimmerman, E. R., Narayan, R., 
	\& McClintock, J. 2005, ApJS, 157, 335

\bibitem[Lightman \& Shapiro(1975)]{lig75} Lightman, A. P., \& Shapiro,
        S. L. 1975, ApJ, 198, L73

\bibitem[Liu et al.(2008)]{liu08} Liu, J., McClintock, J. E., Narayan, R., 
	Davis, S. W., \& Orosz, J. A. 2008, ApJ, 679, L37

\bibitem[Lodato \& Pringle(2006)]{lod06} Lodato, G., \& Pringle,
        J. E. 2006, MNRAS, 368, 1196

\bibitem[Maccarone(2002)]{mac02} Maccarone, T. J. 2002, MNRAS, 336, 1371

\bibitem[Martin et al.(2007)]{mar07} Martin, R. G., Pringle, J. E., \& Tout, 
	C. A. 2007, MNRAS, 381, 1617

\bibitem[Matilsky et al.(1976)]{mat76} Matilsky, T., et al. 1976, ApJ,
        210, L127

\bibitem[McClintock \& Remillard(2006)]{mcc06a} McClintock, J. E., \& 
	Remillard, R. A. 2006, in Compact Stellar X-ray Sources, eds. 
	W. H. G. Lewin \& M. van der Klis (Cambridge: Cambridge Univ. Press), 
	p. 157

\bibitem[McClintock et al.(2008)]{mcc08} McClintock, J. E., Narayan, R., 
	\& Shafee, R. 2008, in Black Holes, eds. M. Livio \& A. 
	Koekemoer (Cambridge Univ. Press), in press (astro-ph/0707.4492)

\bibitem[McClintock et al.(2006)]{mcc06} McClintock, J. E., Shafee, R.,
        Narayan, R., Remillard, R. A., Davis, S. W., \& Li, L.-X. 2006, ApJ,
        652, 518

\bibitem[M\'esz\'aros et al.(1988)]{mes88} M\'esz\'aros, P., Novick, R., 
	Chanan, G. A., Weisskopf, M. C., \& Szentgy\"orgyi, A. 1988, ApJ, 
	324, 1056

\bibitem[Miller(2007)]{mil07} Miller, J. M. 2007, ARA\&A, 45, 441

\bibitem[Miller et al.(2008)]{mil08} Miller, J. M., et al. 2008, ApJ, 679, L113

\bibitem[Mirabel \& Rodr\'iguez(1999)]{mir99} Mirabel, I. F., \&
        Rodr\'iguez, L. F. 1999, ARA\&A, 37, 409

\bibitem[Misner et al.(1973)]{mis73} Misner, C. W., Thorne, K. S., \& Wheeler, 
	J. A. 1973, Gravitation (New York: Freeman)

\bibitem[Narayan \& McClintock(2005)]{nar05} Narayan, R., \& McClintock, J. E. 
	2005, ApJ, 623, 1017

\bibitem[Narayan et al.(2007)]{nar07} Narayan, R., McClintock, J. E., \&
        Shafee, R. 2008, in Astrophysics of Compact Objects, eds. Y. F. Yuan,
        X. D. Li, \& D. Lai (New York: AIP), p. 265

\bibitem[Noble et al.(2008)]{nob08} Noble, S. C., Krolik, J. H., \&
        Hawley, J. F. 2008, ApJ, submitted (astro-ph/0808.3140)

\bibitem[Novick et al.(1972)]{nov72} Novick, R., Weisskopf, M. C.,
        Berthelsdorf, R., Linke, R. \& Wolff, R. S. 1972, ApJ, 174, L1

\bibitem[Novikov \& Thorne(1973)]{nov73} Novikov, I. D., \& Thorne, K. S. 
	1973, in Black Holes, eds. C. DeWitt \& B. S. DeWitt (New York: 
	Gordon and Breach), p. 343

\bibitem[Orosz et al.(2001)]{oro01} Orosz, J. A., et al. 2001, ApJ, 555, 489

\bibitem[Orosz(2003)]{oro03} Orosz, J. A. 2003, in Massive Star Odyssey: 
        From Main Sequence to Supernova, eds. K. van der Hucht, A. Herrero, \& 
        C. Esteban (San Francsisco: ASP), p. 365

\bibitem[Orosz et al.(2007)]{oro07} Orosz, J. A., et al. 2007, Nature, 449, 872

\bibitem[Paczy\'nski(2000)]{pac00}
        Paczy\'nski, B. 2000, astro-ph/0004129

\bibitem[Page \& Thorne(1974)]{pag74} Page, D. N., \& Thorne, K. S. 1974,
	\apj, 191, 499

\bibitem[Popham \& Narayan(1995)]{pop95} Popham, R., \& Narayan, R. 1995, 
	ApJ, 442, 337

\bibitem[Rees(1975)]{ree75} Rees, M. J. 1975, MNRAS, 171, 457

\bibitem[Remillard \& McClintock(2006)]{rem06} Remillard, R. A., \& 
	McClintock, J. E. 2006, ARA\&A, 44, 49

\bibitem[Reynolds \& Fabian(2008)]{rey08} Reynolds, C. S., \& Fabian,
        A. C. 2008, ApJ, 675, 1048

\bibitem[Ross et al.(1999)]{ros99}
	Ross, R. R., Fabian, A. C., \& Young, A. J. 1999, MNRAS, 306, 461

\bibitem[Schnittman(2005)]{sch05} Schnittman, J. D. 2005, ApJ, 621, 940

\bibitem[Shafee et al.(2006)]{sha06} Shafee, R., McClintock, J. E.,
        Narayan, R., Davis, S. W., Li, L.-X., \& Remillard, R. A. 2006, ApJ, 
        636, 113

\bibitem[Shafee et al.(2008a)]{sha08a} Shafee, R., Narayan, R., \& McClintock,
	J. E. 2008a, ApJ, 676, 549

\bibitem[Shafee et al.(2008b)]{sha08b} Shafee, R., McKinney, J. C.,
        Narayan, R., Tchekhovskoy, A., Gammie, C. F., \& McClintock,
        J. E. 2008b, ApJ, submitted (astro-ph/0808.2860)

\bibitem[Shakura \& Sunyaev(1973)]{sha73} Shakura, N. I., \& Sunyaev, R. A.
	1973, A\&A, 24, 337

\bibitem[Shimura \& Takahara(1995)]{shi95} Shimura, T., \& Takahara, F. 1995, 
	ApJ, 445, 780

\bibitem[Silverman \& Filippenko(2008)]{sil08} Silverman, J. M., 
        \& Filippenko, A. V. 2008, ApJ, 678

\bibitem[Stark \& Connors(1977)]{sta77} Stark, R. F., \& Connors, P. A.
	1977, Nature, 266, 429

\bibitem[Swank et al.(2004)]{swa04} Swank, J. H. et al.\ 2004,
        Presentation at the X-ray Polarimetry Workshop held at SLAC, Stanford,
        CA (http://heasarc.gsfc.nasa.gov/docs/heasarc/polar/polar.html)

\bibitem[T\"or\"ok et al.(2005)]{tor05} T\"or\"ok, G., Abramowicz, M. A., 
        Kluz\'niak, W., \& Stuchl\'ik, Z. 2005, A\&A, 436, 1

\bibitem[Tsunemi et al.(1989)]{hir89} Tsunemi, H., Kitamoto, S.,
        Okamura, S., \& Roussel-Dupr\'e, D. 1989, ApJ, 336, L81

\bibitem[van Dam et al.(2006)]{dam06}
	van Dam, M. A., et al. 2006, PASP, 118, 310

\bibitem[Vernet et al.(2007)]{ver07}
        Vernet, J., et al. 2007, ESO Messenger, No.\ 130 (December 2007),
        p.\ 5

\bibitem[Wagoner et al.(2001)]{wag01} Wagoner, R. V., Silbergleit,
        A. S., \& Ortega-Rodríguez, M. 2001, ApJ, 559, 25

\bibitem[Wald(1984)]{wal84} Wald, R. 1984, General Relativity (Chicago: 
	Univ. Chicago Press)

\bibitem[Walker \& Penrose(1970)]{wal70} Walker, M., \& Penrose, R. 1970, 
	Commun. Math. Phys., 18, 265

\bibitem[Watson et al.(1978)]{wat78} Watson, M. G., Ricketts, M. J., \&
        Griffiths, R. E. 1978, ApJ, 221, L69

\bibitem[Weisskopf et al.(1978)]{wei78} Weisskopf, M. C., Silver, E. H.,
        Kestenbaum, H. L., Long, K. S., \& Novick, R. 1978, ApJ, 220, L117

\bibitem[Wilms et al.(2006)]{wil06} Wilms, J., Nowak, M. A.,
Pottschmidt, K., Pooley, G. G., \& Fritz, S. 2006, A\&A, 447, 245

\bibitem[Zahn(1977)]{zah77} Zahn, J.-P. 1977, ApJ, 57, 383


\end{thebibliography}
\end{document}